\documentclass[12pt]{article}
\usepackage[utf8]{inputenc}
\usepackage[margin=1in]{geometry}
\usepackage[style=apa,isbn=false,giveninits=true,uniquename=false,uniquelist=false,maxcitenames=1,backend=biber]{biblatex}
\AtEveryBibitem{\clearfield{note}}
\AtEveryBibitem{\clearfield{url}}
\AtEveryBibitem{\clearfield{publisher}}
\AtEveryBibitem{\clearfield{doi}}
\usepackage{hyperref}
\usepackage[table,xcdraw,dvipsnames]{xcolor}
\hypersetup{colorlinks=true,citecolor=Cyan,linkcolor=Cyan}
\usepackage{multirow}
\usepackage{booktabs}
\usepackage{longtable}
\usepackage{subfiles}
\usepackage{setspace}
\usepackage{lineno}
\usepackage{lscape}
\usepackage{adjustbox}
\usepackage{graphicx}
\usepackage[table]{xcolor}
\usepackage[flushleft]{threeparttable}
\usepackage{latexsym,bm,amsmath,amssymb,amsthm,amsfonts,graphicx}
\providecommand{\keywords}[1]
{
  \small	
  \textbf{\textit{Keywords---}} #1
}

\title{Abstract, keywords and references template}
\author{Author Surname$^{1}$, Someone Else$^{2}$  \\
        \small $^{1}$University A \\
        \small $^{2}$University B \\
}

\def\rv{\text{\rm rv}}

\def\v{\boldsymbol}
\def\E{{\mathbb E}}                         

\newcounter{mntcomm}

\begin{document}

\title{\textbf{Deep Learning Enhanced Realized GARCH}}
\author{Chen Liu\thanks{\textit{Discipline of Business Analytics, The University of Sydney Business School}}
\and Chao Wang\footnotemark[1]
\and Minh-Ngoc Tran\footnotemark[1]
\and Robert Kohn\thanks{\textit{School of Economics, UNSW Business School}}
}
\date{}
\maketitle
\begin{abstract}
We propose a new approach to volatility modeling by combining deep learning (LSTM) and realized volatility measures. This LSTM-enhanced realized GARCH framework incorporates and distills modeling advances from financial econometrics, high frequency trading data and deep learning. Bayesian inference via the Sequential Monte Carlo method is employed for statistical inference and forecasting.
 The new framework can jointly model the returns and realized volatility measures, has an excellent in-sample fit and superior predictive performance compared to several benchmark models, while being able to adapt well to the stylized facts in volatility.
 The performance of the new framework is tested using a wide range of metrics, from marginal likelihood, volatility forecasting, to tail risk forecasting and option pricing.
 We report on a comprehensive empirical study using 31 widely traded stock indices over a time period that includes COVID-19 pandemic.

\end{abstract}
\newpage
\section{Introduction}
Volatility modeling is an active area of research in financial econometrics with implications for risk management, portfolio allocation, and option pricing. The GARCH model (\cite{engle_autoregressive_1982}; \cite{bollerslev_generalized_1986}), and its variants including the exponential GARCH (\cite{nelson_conditional_1991}) and the  GJR model (\cite{glosten_relation_1993}), provide an excellent modeling framework for understanding and forecasting volatility. 
While GARCH and its derivatives work well in many scenarios, when it comes to daily volatility modeling, they predominantly rely on the signal in daily squared returns. This can sometimes fall short in capturing swift intraday volatility fluctuations. 
This problem can be mitigated by incorporating more effective volatility proxies based on high-frequency intraday return data.


In the past two decades, many ex-post estimators of asset return volatility using high-frequency data have been introduced to the literature. Examples include the realized variance of  \cite{andersen_answering_1998}, realized kernel variance of \cite{barndorff-nielsen_designing_2008} and bipower variation of \cite{barndorff-nielsen_power_2004}. These estimators, collectively referred to as realized volatility measures, are more informative than daily squared returns about representing the underlying volatility level, thus providing a better tool for modeling and forecasting volatility. \cite{engle_new_2002} explored the idea of including realized volatility measures in the GARCH model and found that it significantly improves its fit  to return data. Engle's model only uses a realized volatility measure as a deterministic input to the GARCH equation and pays no attention to explaining the variation in realized volatility measures which should be viewed as noisy proxies of the underlying volatility. 
\cite{hansen_realized_2012} developed a complete model, called the realized GARCH (RealGARCH) model, that incorporates a realized measure into a GARCH framework and at the same time provides a measurement equation to explain the realized volatility measure dynamics.
This model has been proven superior empirically \parencite{jiang_modeling_2018,xie_realized_2020,li_efficient_2021}. \cite{hansen_exponential_2016} further extended the RealGARCH model to realized EGARCH which incorporates multiple realized volatility measures of volatility. 

The RealGARCH model expresses the underlying volatility as a {\it linear} combination of several lagged realized measures.
This approach might fall short in grasping the nonlinear relationships and enduring impacts that prior realized measures have on the present intrinsic volatility. 
This issue is similar to the well-known problem of GARCH that has led to the development of  
long-memmory and non-linear GARCH models such as Fractionally Integrated GARCH of \cite{baillie1996fractionally} and Recurrent Conditional Heteroskedasticity (RECH) of \cite{nguyen_recurrent_2022}. This paper aims to extend RealGARCH and address its limitation by leveraging modern deep learning techniques together with recent advancements in Bayesian computation techniques. 

With the advent of deep learning models and advancements in computational power, neural networks (NN) have recently been introduced in volatility modeling in the mainstream econometric literature. NNs are capable of learning complex non-linear functions and capturing long-range dependence in time series data. \cite{liu_novel_2019} and \cite{bucci_realized_2020}  compared the predictive performance of feed-forward neural networks (FNN) and recurrent neural networks (RNN) with traditional econometric approaches, and found that deep learning models generally outperform econometric models on several stock markets.
Some hybrid models, proposed by \cite{hyup_roh_forecasting_2007} and \cite{kim_forecasting_2018}, add a neural network as another layer on top of an econometric model, using the volatility estimates produced by an econometric model as the input to a neural network, which then outputs the final estimate of the volatility. Although these models perform well on specific stock datasets, they are often engineering-oriented and lack interpretability in a financially or economically meaningful way, ignoring important stylized facts such as the leverage and clustering effects commonly observed in financial time series data.
An exception is the recurrent conditional heteroscedastic model (RECH) of \cite{nguyen_recurrent_2022},
that provides a flexible framework for combining deep learning with GARCH-type models.
The RECH model represents the volatility as a sum of two components.
The first component is governed by a GARCH-type model that retains the characteristics and interpretability of  econometric models.
The second component is governed by a RNN that can capture non-linear and long-term serial dependence structure in financial time series.
As demonstrated in \cite{nguyen_recurrent_2022}, RECH retains much of the interpretable characteristics from the GARCH model, while enjoying the modeling flexibility and prediction accuracy from RNN.

This paper proposes a new approach to volatility modeling by combining deep learning and realized volatility measures. Our deep-learning-enhanced realized GARCH framework, under the acronym DeepRGARCH, incorporates and distills modeling advances from financial econometrics, high frequency trading data and deep learning. 
Inspired by RECH, we incorporate the long short-term memory model (LSTM) of \cite{hochreiter_long_1997} into RealGARCH.
The LSTM architecture is one of the most advanced and sophisticated RNN techniques and has proven highly efficient for time series modeling. By incorporating LSTM into RealGARCH, we unlock its modeling power and allow it to be able to capture complex underlying dynamics in financial volatility.   
We compare the performance of the new model with several existing benchmark models on 31 stock market indices. 
We show that the new model substantially improves on previous approaches in both in-sample fit and out-of-sample forecasting.
The code and examples reported in the paper can be found at \url{https://github.com/VBayesLab/DeepRGARCH}.

The rest of the article is organized as follows. Section \ref{sec: model formulation} briefly reviews the model components (the GARCH-type models, realized volatility measures and recurrent neural networks) before describing the proposed DeepRGARCH model. Section \ref{sec:inference} presents the Bayesian inference method for DeepRGARCH and Section \ref{sec:inference} presents the empirical analysis.
Section \ref{sec:conclusion} concludes.
The Appendix provides further implementation and empirical details.

\section{Model Formulation}\label{sec: model formulation}
\subsection{Conditional heteroscedastic models and realized volatility measures}
Let $\v y=\left\{y_t, t=1, \ldots, T\right\}$ be a time series of daily returns. The key term of interest in volatility modeling is the conditional variances, $\sigma_t^2=\operatorname{var}\left(y_t \mid \mathcal{F}_{t-1}\right),$ where $\mathcal{F}_{t-1}$ denotes the $\sigma$-field of information up to and including time $t-1$. 
We assume here that $\E(y_t|\mathcal{F}_{t-1})=0$, but the present method is  easily extended to model the conditional mean $\E(y_t|\mathcal{F}_{t-1})$.
The GARCH model expresses the conditional variance $\sigma_t^2$ as a linear combination of the previous squared returns and conditional variances as an ARMA($p$, $q$) model:
\begin{eqnarray}
    y_{t}&=&\sigma_{t} \epsilon_{t},\quad \epsilon_{t} \overset{\mathrm{i.i.d}}{\sim} \mathcal{N}(0,1) ,\quad t=1,2, \ldots, T\\
    \sigma_{t}^{2}&=&\omega+\sum_{i=1}^{p} \alpha_{i} y_{t-i}^{2}+\sum_{j=1}^{q} \beta_{j} \sigma_{t-j}^{2}, \quad t=p+1, \ldots, T.
\end{eqnarray}
The restriction $\omega>0, \alpha_{i}, \beta_{j} \geq 0, i=1, \ldots, p, j=1, \ldots, q$ is used to ensure positivity of $\sigma_{t}^{2}$, and $\sum_{i=1}^{p} \alpha_{i}+\sum_{j=1}^{q} \beta_{j}<1$ is needed to ensure the stationarity of the time series $y_t$. The errors $\epsilon_t$ are independently and identically distributed as normal distributions with zero mean and unit variance; other distributions for $\epsilon_t$ such as Student's $t$  are also considered in the literature,  e.g., \cite{gerlach_forecasting_2016}. For other GARCH-type models, the reader is referred to \cite{nelson_conditional_1991}, \cite{glosten_relation_1993} and \cite{bollerslev_glossary_2008}.
\par
The GARCH model relies on daily squared returns, which only contain a weak signal of the daily volatility $\sigma_t^2$. It is widely known in the financial econometric literature that high-frequency return data, such as 5-minute data, can be used to estimate daily volatility with high accuracy. 
In the past twenty years, many estimators of daily volatility using high-frequency data were developed and are referred to as ``realized volatility measures" \parencite{andersen_answering_1998,barndorff-nielsen_power_2004,barndorff-nielsen_designing_2008}. 
As realized volatility measures are ex-post, they cannot be directly used for volatility forecasting but they are effective volatility proxies for volatility modeling. \cite{engle_new_2002} is among the first to explore this idea by incorporating the realized volatility measure of 
\cite{andersen_answering_1998} into the GARCH model.
Since then, many volatility models incorporating realized volatility measures have been developed, e.g., \cite{forsberg_bridging_2002}, \cite{engle_multiple_2006}, \cite{corsi_simple_2009}, \cite{shephard_realising_2010}.
The realized GARCH model (RealGARCH) of \cite{hansen_realized_2012} 
\begin{subequations}\label{eq:RealGARCH}
\begin{align}
    y_{t}&=\sigma_{t} \epsilon_{t}, \quad t=1,2, \ldots, T \label{eq:RealGARCH_y}\\
    \sigma^2_{t}&=\omega+\gamma \rv_{t-1}+\beta \sigma^2_{t-1}\label{eq:RealGARCH_sigma}\\
    \rv_{t}&=\xi+\varphi \sigma^2_{t}+\tau\left(\epsilon_{t}\right)+u_{t} \label{eq:RealGARCH_rv}
\end{align}
\end{subequations}
is an important development in this direction. Here $\epsilon_{t}\stackrel{i.i.d}{\sim} N(0,1)$, $u_{t} \stackrel{i.i.d}{\sim} N\left(0, \sigma_{u}^{2}\right)$, $\rv_t$ is a realized volatility measure, and $\tau(\epsilon)$ is regarded as the leverage function and used to capture the leverage effect often observed in volatility. \cite{hansen_realized_2012} set $\tau(\epsilon)=\tau_{1} \epsilon+\tau_{2}\left(\epsilon^{2}-1\right)$. 
An attractive feature of the RealGARCH model is that it contains the measurement equation \eqref{eq:RealGARCH_rv} that
accounts for the variation in the realized volatility measure $\rv_t$. It associates the observed realized volatility measure with the underlying latent volatility, in which the integrated high-frequency variance $\rv_t$ is explained as a linear combination of $\sigma_t^2$ plus a random innovation. 

Our article employs a simple yet effective realized volatility measure, the 5-minute realized variance ($\text{RV}_5$) of \cite{andersen_answering_1998}. Suppose that we observe the asset price at $n$ trading times within a trading day $t$, $t_{j}=t-1+j/n, j=1, \ldots, n$.
Let $\{P(t_j),j=1,...,n\}$ be the observed prices and $r_{t_{j}}=\log P\left(t_{j}\right)-\log P\left(t_{j-1}\right)$ be the log-returns. The RV for the trading day $t$ is defined as
\begin{equation*}\label{eq:rv}
    \rv_t:=\sum_{j=1}^{n} r_{t_{j}}^{2}.
\end{equation*}
It can be shown that \parencite{andersen_answering_1998}, as $n\to\infty$, $\rv_t$ converges in probability to the true latent variance $\sigma_t^2$. 
For $\text{RV}_5$, the return $r_{t_j}$ in the above equation is recorded at 5 minutes frequency.
There are various definitions of realized volatility measures but there is little evidence that any outperform $\text{RV}_5$ as a volatility proxy; see \cite{liu_does_2015} for a detailed comparison of more than 400 realized volatility measures.

\subsection{Recurrent Neural Network}\label{sec:RNN}
RNN is a special class of neural network designed for modeling sequential data. 
Let $\{D_t=(x_t,y_t),t=1,2,...\}$ be the data with $x_t$ the input and $y_t$ the output. 
We use $(x_t,y_t)$ in this section as generic notation, not necessarily applicable to the return data in other sections.
The task is to model the conditional mean $\widehat{y_t}=\E(y_t|x_t,D_{1:t-1})$.
The basic RNN framework is
\begin{subequations}\label{eq:rnn}
\begin{align}
    h_t&=g_h\left(W^{h}\left[h_{t-1}, x_t\right]+b^{h}\right), \label{eq:rnn_a}\\
    \widehat{y_t}&=g_y\left(W^{y} h_t+b^{y}\right). \label{eq:rnn_y}
\end{align}
\end{subequations}
The main feature of the RNN structure is its vector of hidden states $h_t$ which is defined recurrently.
At each time $t$, two information sources are fed into $h_t$: the historical information stored in $h_{t-1}$ and the current information from the input $x_t$.
The functions $g_h$ and $g_y$ are activation functions such as $\text{sig} (z):=1/(1+e^{-z})$, or  $\text{tanh} (z):=(e^{z}-e^{-z})/(e^{z}+e^{-z})$.  
Finally, the $W$ and $b$ are trainable model parameters.

The basic RNN model in \eqref{eq:rnn_a}-\eqref{eq:rnn_y} has some limitations 
in terms of both modeling flexibility and training difficulty.
Many sophisticated RNN structures are proposed to overcome these limitations,
and the LSTM model of \cite{hochreiter_long_1997} stands out as one of the most successful methods.
LSTM uses a gate structure to control the memory in the data. It is written as follows:
\begin{subequations}\label{eq:lstm}
\begin{align}
    g^i_t&=\text{sig}\left(W^{i}\left[h_{t-1}, x_t\right]+b^{i}\right) \label{eq:lstm_gi}\\
    g^f_t&=\text{sig}\left(W^{f}\left[h_{t-1}, x_t\right]+b^{f}\right) \label{eq:lstm_gf}\\
    g^o_t&=\text{sig}\left(W^{o}\left[h_{t-1}, x_t\right]+b^{o}\right) \label{eq:lstm_go}\\
    \tilde{c}_t&=\tanh \left(W^{c}\left[h_{t-1}, x_t\right]+b^{c}\right) \label{eq:lstm_chat}\\
    c_t&=g_t^i \cdot \tilde{c}_t+g_t^f \cdot c_{t-1} \label{eq:lstm_c}\\
    h_{t}&=g_{t}^{o} \cdot \tanh \left(c_{t}\right) \label{eq:lstm_a}\\
    \widehat{y}_{t}&=g_y\left(W^y h_t+b^y\right). \label{eq:lstm_y}
\end{align}
\end{subequations}
Unlike the basic RNN that fully overwrites the memory stored in the hidden states at each step, LSTM can decide to keep, forget or update the memory 
via the memory cell $c_t$ in \eqref{eq:lstm_c}.
This memory cell $c_t$ is updated by partially forgetting the previous memory from $c_{t-1}$ and adding new memory from $\tilde{c}_t$. The extent of forgetting the history and adding new information is controlled by the forget gate $g^f_t$ and input gate $g^i_t$, respectively. Finally, the degree of current memory usage for final output is controlled by the output gate $g^o_t$. See \cite{goodfellow_deep_2016}, for a book length discussion of RNN and deep learning methods in general.

The ability of LSTM to control its memory and quickly adapt to new data patterns makes  it very suitable to model volatility dynamics.
It is well-known that volatility dynamics has long memory as well as a clustering effect
in which the volatility can be highly volatile from period to period \parencite{Benoit1967,Ding:1993,Christensen_Nielsen_long_memory}. Section \ref{sec:DeepRGARCH} describes how to incorporate LSTM into volatility modeling.

\subsection{Deep Learning Enhanced Realized GARCH}\label{sec:DeepRGARCH}
This section presents our DeepRGARCH model,
that incorporates LSTM into RealGARCH for flexible and accurate volatility modeling. Its key motivation is using an additive component
governed by LSTM, that takes the realized measure as its input, to capture complex serial dependence structure in the volatility dynamics.
As intraday realized volatility measures are accurate volatility proxies \parencite{andersen_answering_1998,barndorff-nielsen_designing_2008}, 
feeding them to the LSTM unit further empowers its modeling capacity.  
The DeepRGARCH model of order $p$ and $q$, DeepRGARCH($p,q$), is written as:
\begin{subequations}\label{eq:DeepRGARCH}
\begin{align}
    y_{t}&=\sigma_{t} \epsilon_{t}, \quad t=1,2, \ldots, T\label{eq:DeepRGARCH_yt}\\
    \sigma_{t}^{2}&=g(\omega_t)+\sum_{i=1}^{p} \gamma_{i} \mathrm{rv}_{t-i} + \sum_{j=1}^{q}\beta_{j}\sigma_{t-j}^{2}\label{eq:DeepRGARCH_sigma}\\
    \rv_{t}&=\xi+\varphi \sigma^2_{t}+\tau\left(\epsilon_{t}\right)+u_{t} \label{eq:DeepRGARCH_rv}\\
    \omega_{t}&=\operatorname{LSTM}\left(x_{t}\right)\label{eq:DeepRGARCH_omega} \\
    x_{t}&=\left(\omega_{t-1}, y_{t-1}, \sigma_{t-1}^{2}, \mathrm{rv}_{t-1}\right)\label{eq:DeepRGARCH_x}.
\end{align}
\end{subequations}
Compared to the RealGARCH model, equation \eqref{eq:RealGARCH_sigma}, that only allows a linear and short-term dependence of the true latent conditional variance $\sigma_t^2$ on the realized volatility measure, the DeepRGARCH model is much more flexible.
The latter, via its LSTM component $g(\omega_t)$, allows for both non-linear and long-term dependence that the previous realized volatility measures might have on $\sigma_t^2$. 
As $\sigma_{t-1}^2$ is also an input to the LSTM component, DeepRGARCH also allows for complex serial dependence that the previous volatility might have on $\sigma_t^2$.
We found empirical evidence for the inclusion of $\omega_{t-1}$ and $y_{t-1}$ in the input vector $x_{t}$.
The function $g$ in \eqref{eq:DeepRGARCH_sigma} is a non-negative activation function, applied to the $\omega_t$ to ensure the positive conditional variance. We adopt the RELU function, $g\left(\omega_t\right):=\max \left\{\omega_t, 0\right\}$, for this purpose. 

Inspired by RealGARCH, DeepRGARCH includes the measurement equation \eqref{eq:DeepRGARCH_rv} to account for the variation in the realized volatility measures.
Rather than using the linear structure as in \eqref{eq:DeepRGARCH_rv}, one could easily use an RNN model to explain $\rv_t$ based on $\sigma_t^2$; we have tried this, however, did not observe any meaningful improvement. 
We only include one realized volatility measure in the DeepRGARCH model; however, it is easy to incorporate as many realized volatility measures as possible by using them as additional inputs into $x_t$. We only consider DeepRGARCH(1,1) in this paper, which, for ease of reading and later cross-reference, is expressed as
\begin{subequations}\label{eq:DeepRGARCH(1,1)}
\begin{align}
    y_{t}&=\sigma_{t} \epsilon_{t}, \quad t=1,2, \ldots\\
    \sigma_{t}^{2}&=g(\omega_t)+\gamma\mathrm{rv}_{t-1} + \beta\sigma_{t-1}^{2}\label{eq:DeepRGARCH_sigma1}\\
    \rv_{t}&=\xi+\varphi \sigma^2_{t}+\tau\left(\epsilon_{t}\right)+u_{t} \label{eq:DeepRGARCH_rv1}\\
    \omega_{t}&=\beta_0+\beta_1h_t\label{eq:DeepRGARCH_omega1} \\
    h_{t}&=\operatorname{LSTM}\left(x_{t}\right)\label{eq:DeepRGARCH_h} \\
    x_{t}&=\left(\omega_{t-1}, y_{t-1}, \sigma_{t-1}^{2}, \mathrm{rv}_{t-1}\right)\label{eq:DeepRGARCH_x1}.
\end{align}
\end{subequations}

\section{Bayesian inference for the DeepRGARCH model}\label{sec:inference}
\subsection{The likelihood and prior}
We adopt the Bayesian inference approach for the estimation of DeepRGARCH.
Following \cite{hansen_realized_2012}, we assume Gaussian errors $\epsilon_t \stackrel{i.i.d}{\sim} N(0,1)$ and $u_t \stackrel{i.i.d}{\sim} N(0, \sigma_u)$. Hence, given the observed data $(\v y=\{y_1,...,y_T\},\v{rv}=(\rv_1,...,\rv_T))$ the log-likelihood function of the DeepRGARCH model is:
\begin{equation}
    \ell(\v y,\v{rv}|\theta)=-\frac{1}{2} \sum_{t=1}^T\left[\log (2 \pi)+\log \left(\sigma_t^2\right)+y_t^2 / \sigma_t^2\right] - \frac{1}{2} \sum_{t-1}^T\left[\log (2 \pi)+\log \left(\sigma_{u}^2\right)+u_t^2 / \sigma_u^2\right],
\end{equation}
where $u_t=\rv_t-\xi-\varphi \sigma_t^2-\tau_1 y_t/\sigma_t-\tau_2\left(y_t^2/\sigma_t^2-1\right)$. Recall that the vector of model parameters $\theta$ consists of the GARCH and LSTM parameters. For example, the DeepRGARCH(1,1) model with a single hidden state LSTM has 7 GARCH parameters, $(\beta, \gamma, \xi, \varphi, \tau_1, \tau_2,\sigma_{u})$, and 26 LSTM parameters including $\beta_0$, $\beta_1$ and 24 parameters within the LSTM structure. 

For the prior distributions on the GARCH parameters, we use the commonly used priors in the literature; see, e.g., \cite{gerlach_forecasting_2016}.
The prior $N(0,0.1)$ is used for the LSTM parameters. 

\subsection{Model estimation and prediction}
For Bayesian inference and prediction in volatility modeling, the Sequential Monte Carlo (SMC) method \parencite{neal_annealed_2001,chopin_sequential_2002,del_moral_adaptive_2012} is an effective approach for computing rolling-window volatility forecasts which can effectively sample from non-standard posteriors, while also providing the marginal likelihood estimate as a by-product. The SMC technique uses a set of $M$ weighted particles initially sampled from an easy-to-sample distribution, such as the prior $p(\theta)$, with the particles then traversed through intermediate distributions which eventually become the target distribution. See \cite{gunawan_flexible_2022} for a review of the SMC method.

For in-sample model estimation and inference, we use the likelihood annealing version of SMC that samples from the  sequence of distributions
\begin{equation}\label{eq:likelihood_annealing}
    \pi_k(\theta) \propto p(\theta) p\left(\v y,\v{rv} \mid \theta\right)^{\gamma_k},\;\;k=0,1,\ldots K; 
\end{equation}
here, $0=\gamma_0<\gamma_1<\gamma_2<\ldots<\gamma_K=1$ are called the temperature levels.
Reweighting, resampling, and a Markov transition are the three primary components of the SMC approach. Several methods exist to implement SMC in practice, and we briefly describe  one of them now. The collection of weighted particles $\left\{W_{k-1}^j, \theta_{k-1}^j\right\}_{j=1}^M$ that approximate the intermediate distribution $\pi_{k-1}(\theta)$ is reweighted at the start of iteration $k$ to approximate the target $\pi_k(\theta)$. The efficiency of these weighted particles is measured by the effective sample size (ESS) \parencite{kass_markov_1998}
\begin{equation}
    \mathrm{ESS}=\frac{1}{\sum_{j=1}^M\left(W_t^j\right)^2},
\end{equation}
where $W_t:=(W_t^1, \dots, W_t^M)$ is the weight vector for the $M$ particles at time $t$.
If the ESS is below a threshold, the particles are resampled to obtain equally weighted particles, and a Markov kernel with the invariant distribution $\pi_k(\theta)$ is then applied to refresh these equally weighted particles. Following \cite{del_moral_adaptive_2012}, we adaptively choose the tempering sequence $\gamma_k$ in order to maintain an adequate particle diversity.\par

For out-of-sample rolling-window volatility forecasting which updates the posterior each time a new data observation arrives, we employ the data annealing SMC method that samples from the sequence
\begin{equation}
    \pi_t(\theta) \propto p(\theta)p(\v y_{1:T+t},\v{rv}_{1:T+t}|\theta),\;\; t =1,2,...
\end{equation}

\section{Empirical Analysis}
This section studies the performance of the DeepRGARCH(1,1) model and compares it to GARCH, RealGARCH and RECH
using 31 stock indices including  the Amsterdam Exchange Index (AEX),
the Dow Jones Index (DJI), the Frankfurt Stock Exchange (GDAXI) and the Standard and Poor’s 500 Index (SP500).
The datasets were downloaded from the Realized Library of The Oxford-Man Institute. In the main text, we present results for the above mentioned four representative indices and the average over 31 indices to conserve space; the detailed results for all 31 indices are shown in the appendix. Given the closing prices $\{P_{t},t=0,...,T\}$, we compute the demeaned close-to-close return process as
\begin{equation}
    y_{t}=100\left(\log \frac{P_{t}}{P_{t-1}}-\frac{1}{T} \sum_{i=1}^{T} \log \frac{P_{i}}{P_{i-1}}\right), \quad t=1,2, \ldots, T.
\end{equation}\par
We adopt the 5 mins realized variance \parencite{andersen_modeling_2003} as the realized volatility measure, $\rv_t$, in RealGARCH and DeepRGARCH. As the realized volatility measures ignore the overnight variation of the prices and sometimes the variation in the first few minutes of the trading day when recorded prices may contain large errors \parencite{shephard_realising_2010}, we follow \cite{hansen_forecast_2005} and scale the realized volatility measure as
\begin{equation}\label{eq: resecale rv}
    \widetilde{\sigma}_t^2=\hat{c} \cdot \rv_t \text { where } \hat{c}=\frac{\sum_{i=1}^T y_i^2}{\sum_{i=1}^T \rv_i}, \quad t=1,2, \ldots,
\end{equation}
and use $\tilde{\sigma}_t^2$ as the estimate of the latent conditional variance $ \sigma_t^2$. 
Our sample is from  1 January 2004 to 1 January 2022, including the COVID-19 pandemic period, which we divide it into a first half for training and the second half for out of sample analysis. There are, on average, 2139 trading days for both. See appendix for a detailed data description.\par

\subsection{Parameter estimation and in-sample fit}
As mentioned in Section \ref{sec:inference}, the marginal likelihood estimate is a by-product of SMC, which is useful for in-sample model comparison  using a Bayes factor. Given the training data $(\v y = y_{1:T},\v{rv} = \rv_{1:T})$, let $\widehat{p}\left(\v y,\v{rv}|M_i\right)$ be the marginal likelihood estimate of model $M_i$, $i=1,2$. The Bayes factor of model $M_1$ relative to $M_2$ is 
\begin{equation}
    \text{BF}_{M_{1}, M 2}=\frac{\widehat{p}\left(\v y,\v{rv}|M_1\right)}{\widehat{p}\left(\v y,\v{rv}|M_2\right)}.
\end{equation}
The higher the Bayes factor, the more decisive is the evidence that model $M_1$ is superior to $M_2$ \parencite{kass_bayes_1995}.
Table \ref{tab:para} summarizes the parameter estimation (for the five main parameters) and the Bayes factor (with GARCH used as the baseline model) for the four models, GARCH, RealGARCH, RECH and DeepRGARCH. The last panel reports the results for the average over the 31 indices.

\begin{spacing}{}
\begin{table}[!htbp]
\centering
\begin{threeparttable}
\caption{In-sample analysis: Posterior means of the parameters with the posterior standard deviations (in parentheses).}
\label{tab:para}
\begin{tabular}{ccccccccc}
\toprule
 &  & $\alpha$ & $\beta$ & $\beta_0$ & $\beta_1$ & $\gamma$ & Mar.llik & BF \\
\midrule
\multirow[c]{4}{*}{AEX} & garch & \begin{tabular}[c]{@{}c@{}}0.105\\ {\scriptsize(0.012)}\end{tabular} & \begin{tabular}[c]{@{}c@{}}0.884\\ {\scriptsize(0.013)}\end{tabular} & - & - & - & \begin{tabular}[c]{@{}c@{}}$-3450.8$\\ {\scriptsize(0.183)}\end{tabular} & 1 \\
 & rech & \begin{tabular}[c]{@{}c@{}}0.038\\ {\scriptsize(0.014)}\end{tabular} & \begin{tabular}[c]{@{}c@{}}0.741\\ {\scriptsize(0.061)}\end{tabular} & \begin{tabular}[c]{@{}c@{}}0.049\\ {\scriptsize(0.019)}\end{tabular} & \begin{tabular}[c]{@{}c@{}}0.907\\ {\scriptsize(0.182)}\end{tabular} & - & \begin{tabular}[c]{@{}c@{}}$-3400.6$\\ {\scriptsize(0.280)}\end{tabular} & $e^{50}$ \\
 & realgarch & - & \begin{tabular}[c]{@{}c@{}}0.549\\ {\scriptsize(0.028)}\end{tabular} & - & - & \begin{tabular}[c]{@{}c@{}}0.371\\ {\scriptsize(0.028)}\end{tabular} & \begin{tabular}[c]{@{}c@{}}$-3376.3$\\ {\scriptsize(0.018)}\end{tabular} & $e^{75}$ \\
 & deeprgarch  & - & \begin{tabular}[c]{@{}c@{}}0.527\\ {\scriptsize(0.021)}\end{tabular} & \begin{tabular}[c]{@{}c@{}}0.112\\ {\scriptsize(0.022)}\end{tabular} & \begin{tabular}[c]{@{}c@{}}0.801\\ {\scriptsize(0.081)}\end{tabular} & \begin{tabular}[c]{@{}c@{}}0.341\\ {\scriptsize(0.015)}\end{tabular} & \begin{tabular}[c]{@{}c@{}}$-3370.5$\\ {\scriptsize(0.220)}\end{tabular} & $\mathbf{e^{80}}$ \\
\cline{1-9}
\multirow[c]{4}{*}{DJI} & garch & \begin{tabular}[c]{@{}c@{}}0.094\\ {\scriptsize(0.011)}\end{tabular} & \begin{tabular}[c]{@{}c@{}}0.890\\ {\scriptsize(0.012)}\end{tabular} & - & - & - & \begin{tabular}[c]{@{}c@{}}$-3027.2$\\ {\scriptsize(0.100)}\end{tabular} & 1 \\
 & rech & \begin{tabular}[c]{@{}c@{}}0.049\\ {\scriptsize(0.011)}\end{tabular} & \begin{tabular}[c]{@{}c@{}}0.636\\ {\scriptsize(0.063)}\end{tabular} & \begin{tabular}[c]{@{}c@{}}0.063\\ {\scriptsize(0.019)}\end{tabular} & \begin{tabular}[c]{@{}c@{}}0.912\\ {\scriptsize(0.182)}\end{tabular} & - & \begin{tabular}[c]{@{}c@{}}$-2991.0$\\ {\scriptsize(0.785)}\end{tabular} & $e^{36}$ \\
 & realgarch & - & \begin{tabular}[c]{@{}c@{}}0.634\\ {\scriptsize(0.024)}\end{tabular} & - & - & \begin{tabular}[c]{@{}c@{}}0.318\\ {\scriptsize(0.027)}\end{tabular} & \begin{tabular}[c]{@{}c@{}}$-2968.4$\\ {\scriptsize(0.075)}\end{tabular} & $e^{59}$ \\
 & deeprgarch  & - & \begin{tabular}[c]{@{}c@{}}0.840\\ {\scriptsize(0.014)}\end{tabular} & \begin{tabular}[c]{@{}c@{}}1.413\\ {\scriptsize(0.148)}\end{tabular} & \begin{tabular}[c]{@{}c@{}}11.827\\ {\scriptsize(0.577)}\end{tabular} & \begin{tabular}[c]{@{}c@{}}0.005\\ {\scriptsize(0.004)}\end{tabular} & \begin{tabular}[c]{@{}c@{}}$-2962.9$\\ {\scriptsize(0.421)}\end{tabular} & $\mathbf{e^{64}}$ \\
\cline{1-9}
\multirow[c]{4}{*}{GDAXI} & garch & \begin{tabular}[c]{@{}c@{}}0.098\\ {\scriptsize(0.013)}\end{tabular} & \begin{tabular}[c]{@{}c@{}}0.889\\ {\scriptsize(0.014)}\end{tabular} & - & - & - & \begin{tabular}[c]{@{}c@{}}$-3615.5$\\ {\scriptsize(0.107)}\end{tabular} & 1 \\
 & rech & \begin{tabular}[c]{@{}c@{}}0.049\\ {\scriptsize(0.013)}\end{tabular} & \begin{tabular}[c]{@{}c@{}}0.666\\ {\scriptsize(0.071)}\end{tabular} & \begin{tabular}[c]{@{}c@{}}0.079\\ {\scriptsize(0.036)}\end{tabular} & \begin{tabular}[c]{@{}c@{}}1.000\\ {\scriptsize(0.206)}\end{tabular} & - & \begin{tabular}[c]{@{}c@{}}$-3570.0$\\ {\scriptsize(0.349)}\end{tabular} & $e^{46}$ \\
 & realgarch & - & \begin{tabular}[c]{@{}c@{}}0.439\\ {\scriptsize(0.025)}\end{tabular} & - & - & \begin{tabular}[c]{@{}c@{}}0.452\\ {\scriptsize(0.031)}\end{tabular} & \begin{tabular}[c]{@{}c@{}}$-3534.8$\\ {\scriptsize(0.055)}\end{tabular} & $e^{81}$ \\
 & deeprgarch  & - & \begin{tabular}[c]{@{}c@{}}0.414\\ {\scriptsize(0.029)}\end{tabular} & \begin{tabular}[c]{@{}c@{}}0.126\\ {\scriptsize(0.084)}\end{tabular} & \begin{tabular}[c]{@{}c@{}}5.561\\ {\scriptsize(0.409)}\end{tabular} & \begin{tabular}[c]{@{}c@{}}0.307\\ {\scriptsize(0.022)}\end{tabular} & \begin{tabular}[c]{@{}c@{}}$-3523.6$\\ {\scriptsize(0.141)}\end{tabular} & $\mathbf{e^{92}}$ \\
\cline{1-9}
\multirow[c]{4}{*}{SPX} & garch & \begin{tabular}[c]{@{}c@{}}0.091\\ {\scriptsize(0.010)}\end{tabular} & \begin{tabular}[c]{@{}c@{}}0.895\\ {\scriptsize(0.011)}\end{tabular} & - & - & - & \begin{tabular}[c]{@{}c@{}}$-3177.1$\\ {\scriptsize(0.158)}\end{tabular} & 1 \\
 & rech & \begin{tabular}[c]{@{}c@{}}0.049\\ {\scriptsize(0.010)}\end{tabular} & \begin{tabular}[c]{@{}c@{}}0.657\\ {\scriptsize(0.059)}\end{tabular} & \begin{tabular}[c]{@{}c@{}}0.071\\ {\scriptsize(0.020)}\end{tabular} & \begin{tabular}[c]{@{}c@{}}0.871\\ {\scriptsize(0.193)}\end{tabular} & - & \begin{tabular}[c]{@{}c@{}}$-3150.4$\\ {\scriptsize(0.491)}\end{tabular} & $e^{27}$ \\
 & realgarch & - & \begin{tabular}[c]{@{}c@{}}0.633\\ {\scriptsize(0.022)}\end{tabular} & - & - & \begin{tabular}[c]{@{}c@{}}0.317\\ {\scriptsize(0.025)}\end{tabular} & \begin{tabular}[c]{@{}c@{}}$-3100.1$\\ {\scriptsize(0.078)}\end{tabular} & $\mathbf{e^{77}}$ \\
 & deeprgarch  & - & \begin{tabular}[c]{@{}c@{}}0.831\\ {\scriptsize(0.016)}\end{tabular} & \begin{tabular}[c]{@{}c@{}}1.549\\ {\scriptsize(0.160)}\end{tabular} & \begin{tabular}[c]{@{}c@{}}11.432\\ {\scriptsize(0.660)}\end{tabular} & \begin{tabular}[c]{@{}c@{}}0.007\\ {\scriptsize(0.005)}\end{tabular} & \begin{tabular}[c]{@{}c@{}}$-3104.7$\\ {\scriptsize(0.205)}\end{tabular} & $e^{72}$ \\
\cline{1-9}
\multirow[c]{4}{*}{Mean} & garch & \begin{tabular}[c]{@{}c@{}}0.102\\ {\scriptsize(0.013)}\end{tabular} & \begin{tabular}[c]{@{}c@{}}0.882\\ {\scriptsize(0.015)}\end{tabular} & - & - & - & \begin{tabular}[c]{@{}c@{}}$-3411.9$\\ {\scriptsize(0.132)}\end{tabular} & 1 \\
 & rech & \begin{tabular}[c]{@{}c@{}}0.059\\ {\scriptsize(0.014)}\end{tabular} & \begin{tabular}[c]{@{}c@{}}0.729\\ {\scriptsize(0.057)}\end{tabular} & \begin{tabular}[c]{@{}c@{}}0.071\\ {\scriptsize(0.032)}\end{tabular} & \begin{tabular}[c]{@{}c@{}}0.850\\ {\scriptsize(0.193)}\end{tabular} & - & \begin{tabular}[c]{@{}c@{}}$-3384.3$\\ {\scriptsize(0.226)}\end{tabular} & $e^{28}$ \\
 & realgarch & - & \begin{tabular}[c]{@{}c@{}}0.628\\ {\scriptsize(0.024)}\end{tabular} & - & - & \begin{tabular}[c]{@{}c@{}}0.306\\ {\scriptsize(0.025)}\end{tabular} & \begin{tabular}[c]{@{}c@{}}$-3368.4$\\ {\scriptsize(0.214)}\end{tabular} & $e^{44}$ \\
 & deeprgarch  & - & \begin{tabular}[c]{@{}c@{}}0.667\\ {\scriptsize(0.034)}\end{tabular} & \begin{tabular}[c]{@{}c@{}}0.613\\ {\scriptsize(0.118)}\end{tabular} & \begin{tabular}[c]{@{}c@{}}5.203\\ {\scriptsize(0.453)}\end{tabular} & \begin{tabular}[c]{@{}c@{}}0.168\\ {\scriptsize(0.021)}\end{tabular} & \begin{tabular}[c]{@{}c@{}}$-3354.0$\\ {\scriptsize(0.472)}\end{tabular} & $\mathbf{e^{58}}$ \\
\cline{1-9}
\bottomrule
\end{tabular}
\begin{tablenotes}
  \item \footnotesize \emph{Note:} The last two columns show the natural logarithms of the estimated marginal likelihood (Mar.llik) with Monte Carlo standard errors (in parentheses) across 6 different runs of SMC and the Bayes factors (BF) that uses GARCH as the baseline.
\end{tablenotes}
\end{threeparttable}
\end{table}
\end{spacing}

We draw the following conclusions from the estimation results. First, the marginal likelihood estimates show that the DeepRGARCH model fits the index datasets better than the competing models for 28 of 31 indices. On average, the Bayes factors of the DeepRGARCH model compared with the GARCH, RECH, and RealGARCH models are roughly $e^{58}$, $e^{30}$ and $e^{14}$, respectively, which according to Jeffery's scale for interpreting the Bayes factor \parencite{jeffreys_theory_1998}, decisively support the DeepRGARCH model.
Second, the estimated posterior mean of the parameter $\beta_1$ in the DeepRGARCH model is significantly different from zero in all cases, providing evidence of the volatility effects rather than linearity that the previous conditional variance $\sigma_{t-1}^2$ and realized volatility measure $\rv_{t-1}$ have on $\sigma_t^2$. This also suggests that the LSTM component $g(\omega_t)$ in DeepRGARCH can detect these effects effectively. Additionally, the estimated value of the parameter $\gamma$ (the coefficient concerning the realized volatility measure) in DeepRGARCH is always smaller than that in RealGARCH. This is perhaps because the effect of the realized volatility measure $\rv_{t-1}$ on the underlying volatility $\sigma_t^2$ is well captured by the LSTM component, which, unlike RealGARCH, allows non-linear dependence of $\sigma_t^2$ on $\rv_{t-1}$.

\subsection{Volatility forecast error compared to ex-post realized volatility measures}
We now test the predictive performance of the DeepRGARCH model for volatility forecasting.
The forecast performance is measured by mean squared error (MSE) and mean absolute deviation (MAD) computed on test data $D_{test}$ of size $T_{test}$
\begin{eqnarray}
\text{MSE}&=&T_{test}^{-1}\sum_{D_{test}} (\widehat{\sigma}_t-\widetilde{\sigma_t})^2\\
\text{MAE}&=&T_{test}^{-1}\sum_{D_{test}} |\widehat{\sigma}_t-\widetilde{\sigma_t}|  
\end{eqnarray}
where $\widehat{\sigma}_t$ is the one-step-ahead rolling-window forecast of the latent $\sigma_t$
and $\widetilde{\sigma_t}$ is the squared root of an ex-post realized volatility measure after rescaling as in \eqref{eq: resecale rv}.     
We use five ex-post volatility proxies, the Realized Variance (RV), Bipower Variation (BV), Median Realized Volatility (MedRV), Realized Kernel Variance with the Non-flat Parzen kernel (RK-Parzen) and the Tukey-Hanning kernel (RSV). See \cite{shephard_realising_2010} for details about the Realized Library.

Tables \ref{tab:mse} and \ref{tab:mad} summarize the forecast performance of the four models.
The last column in each table shows the number of times a model has the best predictive score across the five volatility proxies.
The last panel in each table shows the average scores over the 31 indices.
Using the five ex-post realized volatility measures as the ground truth, the results show that the DeepRGARCH model performs the best in terms of volatility forecasting.
\begin{spacing}{}
\begin{table}[!htbp]
\centering
\begin{threeparttable}
\caption{Forecast performance: MSE for different realized volatility measures.}
\label{tab:mse}
\begin{tabular}{cccccccc}
\toprule
 &  & RV5 & BV & MedRV & RK-Parzen & RSV & Count \\
\midrule
\multirow[c]{4}{*}{AEX} & garch & 0.151 & 0.142 & 0.141 & 0.209 & 0.190 & 0.0 \\
 & rech & 0.143 & 0.131 & 0.126 & 0.188 & 0.180 & 0.0 \\
 & realgarch & 0.121 & 0.109 & 0.102 & 0.173 & 0.158 & 0.0 \\
 & deeprgarch  & \bfseries 0.116 & \bfseries 0.103 & \bfseries 0.094 & \bfseries 0.157 & \bfseries 0.152 & \bfseries 5.0 \\
\cline{1-8}
\multirow[c]{4}{*}{DJI} & garch & 0.203 & 0.174 & 0.177 & 0.202 & 0.274 & 0.0 \\
 & rech & 0.194 & 0.176 & 0.176 & 0.193 & 0.266 & 0.0 \\
 & realgarch & 0.121 & 0.124 & 0.127 & 0.119 & 0.196 & 0.0 \\
 & deeprgarch  & \bfseries 0.112 & \bfseries 0.122 & \bfseries 0.118 & \bfseries 0.109 & \bfseries 0.184 & \bfseries 5.0 \\
\cline{1-8}
\multirow[c]{4}{*}{GDAXI} & garch & 0.180 & 0.174 & 0.182 & 0.203 & 0.221 & 0.0 \\
 & rech & 0.159 & 0.154 & 0.160 & 0.180 & 0.200 & 0.0 \\
 & realgarch & 0.111 & 0.110 & 0.114 & 0.140 & 0.152 & 0.0 \\
 & deeprgarch  & \bfseries 0.106 & \bfseries 0.104 & \bfseries 0.105 & \bfseries 0.132 & \bfseries 0.146 & \bfseries 5.0 \\
\cline{1-8}
\multirow[c]{4}{*}{SPX} & garch & 0.183 & 0.166 & 0.180 & 0.182 & 0.243 & 0.0 \\
 & rech & 0.174 & 0.163 & 0.172 & 0.172 & 0.237 & 0.0 \\
 & realgarch & 0.117 & \bfseries 0.119 & 0.129 & 0.119 & 0.182 & 1.0 \\
 & deeprgarch  & \bfseries 0.112 & 0.121 & \bfseries 0.125 & \bfseries 0.113 & \bfseries 0.176 & \bfseries 4.0 \\
\cline{1-8}
\multirow[c]{4}{*}{Mean} & garch & 0.196 & 0.174 & 0.184 & 0.243 & 0.267 & 0.0 \\
 & rech & 0.184 & 0.163 & 0.172 & 0.227 & 0.257 & 0.3 \\
 & realgarch & 0.153 & 0.138 & 0.151 & 0.204 & 0.225 & 1.2 \\
 & deeprgarch  & \bfseries 0.147 & \bfseries 0.133 & \bfseries 0.146 & \bfseries 0.192 & \bfseries 0.219 & \bfseries 3.5 \\
\cline{1-8}
\bottomrule
\end{tabular}
\begin{tablenotes}
  \item \footnotesize \emph{Note:} The last column reports the number of times a model has the lowest MSE among the five realized volatility measures. The bottom panel reports the average across the 31 indices. The bold numbers indicate the best scores. 
\end{tablenotes}
\end{threeparttable}
\end{table}
\end{spacing}

\begin{spacing}{}
\begin{table}[!htbp]
\centering
\begin{threeparttable}
\caption{Forecast performance: MAD for different realized volatility measures.}
\label{tab:mad}
\begin{tabular}{cccccccc}
\toprule
 &  & RV5 & BV & MedRV & RK-Parzen & RSV & Count \\
\midrule
\multirow[c]{4}{*}{AEX} & garch & 0.259 & 0.256 & 0.266 & 0.332 & 0.298 & 0.0 \\
 & rech & 0.237 & 0.233 & 0.242 & 0.300 & 0.276 & 0.0 \\
 & realgarch & 0.218 & 0.211 & 0.219 & 0.294 & 0.260 & 0.0 \\
 & deeprgarch  & \bfseries 0.209 & \bfseries 0.202 & \bfseries 0.209 & \bfseries 0.274 & \bfseries 0.249 & \bfseries 5.0 \\
\cline{1-8}
\multirow[c]{4}{*}{DJI} & garch & 0.291 & 0.273 & 0.301 & 0.308 & 0.344 & 0.0 \\
 & rech & 0.267 & 0.253 & 0.278 & 0.281 & 0.323 & 0.0 \\
 & realgarch & 0.218 & 0.214 & 0.239 & 0.237 & 0.279 & 0.0 \\
 & deeprgarch  & \bfseries 0.208 & \bfseries 0.204 & \bfseries 0.226 & \bfseries 0.223 & \bfseries 0.272 & \bfseries 5.0 \\
\cline{1-8}
\multirow[c]{4}{*}{GDAXI} & garch & 0.317 & 0.310 & 0.320 & 0.343 & 0.356 & 0.0 \\
 & rech & 0.297 & 0.290 & 0.297 & 0.324 & 0.337 & 0.0 \\
 & realgarch & 0.237 & 0.233 & 0.242 & 0.274 & 0.284 & 0.0 \\
 & deeprgarch  & \bfseries 0.233 & \bfseries 0.229 & \bfseries 0.237 & \bfseries 0.266 & \bfseries 0.277 & \bfseries 5.0 \\
\cline{1-8}
\multirow[c]{4}{*}{SPX} & garch & 0.295 & 0.286 & 0.312 & 0.305 & 0.340 & 0.0 \\
 & rech & 0.265 & 0.262 & 0.285 & 0.275 & 0.315 & 0.0 \\
 & realgarch & 0.222 & 0.225 & 0.247 & 0.240 & 0.277 & 0.0 \\
 & deeprgarch  & \bfseries 0.211 & \bfseries 0.217 & \bfseries 0.237 & \bfseries 0.226 & \bfseries 0.270 & \bfseries 5.0 \\
\cline{1-8}
\multirow[c]{4}{*}{Mean} & garch & 0.289 & 0.278 & 0.293 & 0.338 & 0.337 & 0.0 \\
 & rech & 0.278 & 0.267 & 0.282 & 0.323 & 0.327 & 0.2 \\
 & realgarch & 0.248 & 0.238 & 0.258 & 0.305 & 0.302 & 0.7 \\
 & deeprgarch  & \bfseries 0.241 & \bfseries 0.233 & \bfseries 0.254 & \bfseries 0.292 & \bfseries 0.294 & \bfseries 4.1 \\
\cline{1-8}
\bottomrule
\end{tabular}
\begin{tablenotes}
  \item \footnotesize \emph{Note:}  The last column reports the number of times a model has the lowest MAD among the five realized volatility measures. The last panel reports the average across th 31 indices. The bold numbers indicate the best scores. 
\end{tablenotes}
\end{threeparttable}
\end{table}
\end{spacing}

\subsection{Fitness to return series and tail risk forecast}
An attractive feature of GARCH-type models including DeepRGARCH is that they 
 model both the volatility  and return 
 processes. This feature is crucial to risk management which is one of the most important applications of volatility modeling.
For risk management, the key task is to forecast the Value at Risk (VaR) and Expected Shortfall (ES). 
An $\alpha$-level VaR is the $\alpha$-level quantile of the distribution of the return,
and the $\alpha$-level ES is the conditional expectation of return values that exceed the corresponding $\alpha$-level VaR.
Both VaR and ES are used as the two key risk measures in financial regulation 
and recommended by the Basel Accord.

To evaluate the quality of the VaR forecast, we adopt the standard quantile loss function 
\begin{equation}
    \text{Qloss} := \sum_{y_t \in D_{test}}\left(\alpha-I\left(y_t<Q_t^\alpha\right)\right)\left(y_t-Q_t^\alpha\right),
\end{equation}
where $Q_t^\alpha$ is the forecast of $\alpha$-level VaR of $y_t$ \parencite{koenker_regression_1978}.
We note that, given the volatility $\sigma_t$ and the normality assumption of the random shock $\epsilon_t$ in \eqref{eq:DeepRGARCH_yt}, it is straightforward to compute $Q_t^\alpha$.
The quantile loss function is strictly consistent \parencite{fissler_higher_2016}, i.e., the expected loss is lowest at the true quantile series. The most accurate VaR forecasting model should therefore minimize the quantile loss function.

There is no strictly consistent loss function for ES;
however, \cite{fissler_higher_2016} found that ES and
VaR are jointly elicitable, i.e., there is a class of strictly consistent loss functions for evaluating
VaR and ES forecasts jointly. \cite{taylor_forecasting_2019} showed that the negative logarithm of the likelihood function built from the Asymmetric Laplace (AL) distribution is strictly consistent for VaR and ES considered jointly and fits into the class developed by \cite{fissler_higher_2016}. This AL based joint loss function is given as
\begin{equation}
    \text{JointLoss} := \frac{1}{T_{\text {test }}}\sum_{D_{\text {test}}}\left(-\log \left(\frac{\alpha-1}{\mathrm{ES}_t^\alpha}\right)-\frac{\left(y_t-Q_t^\alpha\right)\left(\alpha-I\left(y_t \leq Q_t^\alpha\right)\right)}{\alpha \mathrm{ES}_t^\alpha}\right)
\end{equation}
with $\mathrm{ES}_t^\alpha$ the forecast of $\alpha$-level ES of $y_t$.

To measure the fit of a volatility model to the return series,
we also consider the Partial Predictive Score (PPS), which is one of the most commonly used metrics for evaluating predictive performance in statistical modeling. It measures the negative log-likelihood of observing the return series based on our volatility forecast. The model with the smallest PPS is preferred. The PPS score is defined as
\begin{equation}
    \text{PPS}:=-\frac{1}{T_{\text {test }}} \sum_{D_{\text {test }}} p\left(y_{t} \mid y_{1: t-1}\right).
\end{equation}
Table \ref{tab:risk} reports the quantile loss, joint loss for 1\% and 5\% VaR and ES forecast, and PPS score. The Count panel reports the number of indices where a model achieves the best predictive score. The results show that for most of the indices, the DeepRGARCH model produces the best VaR and ES forecasts. DeepRGARCH is also superior to its competitors in terms of overall fit to the return series. It reports the lowest PPS on most returns series and on average. Additionally, we find that for most of the metrics, the RECH model performs the best on more indices than RealGARCH. In contrast, in the previous section, RealGARCH dominates RECH regarding forecast error compared ex-post proxies. This suggests that ranking conditional volatility forecasts by ex-post proxies can sometimes lead to undesirable outcomes, and we should also evaluate predictive performance by economic loss.  

\begin{spacing}{}
\begin{table}[!htbp]
\centering
\begin{threeparttable}
\caption{Forecast performance: Tail risk forecast and Partial Predictive Score.}
\label{tab:risk}
\begin{tabular}{ccccccc}
\toprule
 &  & Qloss\_1\% & JointLoss\_1\% & Qloss\_5\% & JointLoss\_5\% & PPS \\
\midrule
\multirow[c]{4}{*}{AEX} & garch & 92.549 & 2.462 & 282.507 & 1.909 & 1.342 \\
 & rech & 88.654 & 2.404 & 270.841 & 1.851 & 1.307 \\
 & realgarch & 86.838 & 2.317 & 273.663 & 1.847 & 1.298 \\
 & deeprgarch  & \bfseries 84.982 & \bfseries 2.279 & \bfseries 267.902 & \bfseries 1.815 & \bfseries 1.288 \\
\cline{1-7}
\multirow[c]{4}{*}{DJI} & garch & \bfseries 80.691 & 2.340 & 250.738 & 1.767 & 1.175 \\
 & rech & 81.502 & 2.243 & 247.145 & 1.706 & 1.135 \\
 & realgarch & 82.620 & 2.299 & 242.673 & 1.712 & 1.142 \\
 & deeprgarch  & 81.193 & \bfseries 2.236 & \bfseries 242.049 & \bfseries 1.689 & \bfseries 1.134 \\
\cline{1-7}
\multirow[c]{4}{*}{GDAXI} & garch & 100.503 & 2.514 & 316.005 & 2.034 & 1.484 \\
 & rech & \bfseries 97.312 & \bfseries 2.480 & \bfseries 305.339 & \bfseries 1.991 & \bfseries 1.453 \\
 & realgarch & 103.330 & 2.577 & 314.835 & 2.039 & 1.460 \\
 & deeprgarch  & 100.372 & 2.557 & 309.042 & 2.016 & 1.453 \\
\cline{1-7}
\multirow[c]{4}{*}{SPX} & garch & 83.164 & 2.424 & 253.502 & 1.803 & 1.192 \\
 & rech & 81.716 & 2.327 & 247.041 & 1.735 & 1.149 \\
 & realgarch & \bfseries 80.571 & 2.311 & 244.235 & 1.728 & 1.136 \\
 & deeprgarch  & 81.334 & \bfseries 2.287 & \bfseries 243.364 & \bfseries 1.714 & \bfseries 1.135 \\
\cline{1-7}
\multirow[c]{4}{*}{Mean} & garch & 78.545 & 2.262 & 260.099 & 1.859 & 1.355 \\
 & rech & 76.549 & 2.224 & 253.029 & 1.825 & 1.339 \\
 & realgarch & 79.518 & 2.266 & 258.329 & 1.853 & 1.348 \\
 & deeprgarch  & \bfseries 76.141 & \bfseries 2.223 & \bfseries 248.735 & \bfseries 1.812 & \bfseries 1.336 \\
\cline{1-7}
\multirow[c]{4}{*}{Count} & garch & 1 & 3 & 0 & 0 & 0 \\
 & rech & 9 & 7 & 6 & 7 & 11 \\
 & realgarch & 4 & 3 & 3 & 1 & 4 \\
 & deeprgarch  & \bfseries 17 & \bfseries 18 & \bfseries 22 & \bfseries 23 & \bfseries 16 \\
\cline{1-7}
\bottomrule
\end{tabular}
\begin{tablenotes}
  \item \footnotesize \emph{Note:} Qloss\_1\%, JointLoss\_1\%, Qloss\_5\%, JointLoss\_5\% are the quantile loss and jointloss at 1\% and 5\% respectively. The last two panels report the average scores and the number of times a model has the best predictive scores across the 31 indices. 
\end{tablenotes}
\end{threeparttable}
\end{table}
\end{spacing}

\subsection{Simulated option trading}
Apart from risk management, options trading is also one of the most attractive applications of volatility forecasting. In a theoretical pricing model, volatility is the most difficult input for traders to predict among all the inputs required for option evaluation. At the same time, volatility often plays the most crucial role in actual trading decisions. Consider the inputs of a Black-Scholes model for European options:
\begin{enumerate}
	\item The current price of the underlying security
	\item The option's exercise price
	\item The expiration time
	\item The risk-free interest rate of the life of the option
	\item The volatility of the underlying contract. 
\end{enumerate}
Volatility is the only unknown input here; hence the profitability of an options trader is greatly affected by their ability to forecast volatility. This section examines model performance based on its ability to price options correctly. We follow \cite{engle_valuation_1990} to design a hypothetical option market where each agent uses their volatility forecast and the Black-Scholes model to price options and trade with competing agents. The experiment is organized as follows:
\begin{enumerate}
	\item An agent in the experiment trades on the options of \$1 share of the S\&P500 index with an at-the-money exercise price (1\$) and 1-day expiration. The risk-free interest rate is set to zero. 
	\item Each agent $M$ determines their call options price 
	\begin{equation}
		P_{t,M}=2 \Phi\left(\frac{1}{2} \sigma_{t,M}\right)-1
	\end{equation}
 given the volatility forecast $\sigma_{t, M}^2$ and the Black-Scholes formula, 
	with $\Phi$ the standard normal cumulative distribution function.
	\item The pair-wise trading then takes place between agents $M_1$ and $M_2$ at their predicted mid-price $P_t$, 
	\begin{equation}
		P_t = (P_{t,M_1}+P_{t,M_2})/2.
	\end{equation}
	Each agent either buys or sells a straddle (a combination of put and call options) and uses its variance forecast to determine the hedge ratio, $\delta$. In our case, $\delta_\text{straddle}=1-2 \Phi\left(\frac{1}{2} \sigma_t\right)$. The intuition is that the agent with the higher volatility forecast will believe the straddle is underpriced from $P_t$, thus buying the straddle from its counterpart and vice versa. 
	\item For each pair-wise trade, the daily profit of buying a straddle is then calculated as
	\begin{equation}
		\left|r_t\right|-2 P_t+r_t\left(1-2 \Phi\left(\frac{1}{2} \sigma_{t,M}\right)\right),
	\end{equation}
	and the daily profit of selling a straddle is
	\begin{equation}
		2 P_t-\left|r_t\right|-r_t\left(1-2 \Phi\left(\frac{1}{2} \sigma_{t,M}\right)\right).
	\end{equation}
	With a total of $k$ agents, each agent conducts $k-1$ trades per day. The daily sum of the trading profit is then divided by $k-1$ and averaged throughout the testing period.
\end{enumerate}\par 
Table~\ref{tab:opt} reports the daily, annual profit (in cents) and Sharpe ratio of the options agents that use GARCH, RealGARCH, RECH and DeepRGARCH as their forecast models. The simulations are repeated on 31 stocks indices and the following four scenarios:
\begin{enumerate}
	\item All agents trade in the market.
	\item Only RealGARCH, RECH, and DeepRGARCH agents trade in the market.
	\item Only RealGARCH and DeepRGARCH agents trade against each other. 
	\item Only RECH and DeepRGARCH agents trade against each other. 
\end{enumerate}
In scenario (1), where all agents are trading against each other, the DeepRGARCH agent generates the highest profits and Sharpe ratio in most stock indices on average. Scenarios (2), (3) to  (4) further illustrate the consistent profitability of the DeepRGARCH agent against its two direct ancestors, RealGARCH and RECH, in our hypothetical market. A surprising finding is that the RealGARCH performs almost as badly as the GARCH model. 
\begin{spacing}{}
\begin{table}[!htbp]
\centering
\begin{threeparttable}
\caption{Forecast performance: Annualized return (Ret.) and Sharpe ratio (Sharpe) of option trading simulation}
\label{tab:opt}
\begin{tabular}{rrrrrrrrrr}
\toprule
 &  & \multicolumn{2}{r}{Scenario1} & \multicolumn{2}{r}{Scenario2} & \multicolumn{2}{r}{Scenario3} & \multicolumn{2}{r}{Scenario4} \\
 &  & Ret. & Sharpe & Ret. & Sharpe & Ret. & Sharpe & Ret. & Sharpe \\
\midrule
\multirow[c]{4}{*}{AEX} & garch & $ -22.4 $ & $ -3.2 $ & - & - & - & - & - & - \\
 & rech & 5.9 & 0.5 & $ -4.1 $ & $ -0.6 $ & $ -9.5 $ & $ -1.1 $ & - & - \\
 & realgarch & 1.9 & 0.0 & $ -7.4 $ & $ -1.0 $ & - & - & $ -12.4 $ & $ -1.4 $ \\
 & deeprgarch  & \bfseries 17.8 & \bfseries 1.9 & \bfseries 11.3 & \bfseries 1.2 & \bfseries 9.2 & \bfseries 0.7 & \bfseries 12.9 & \bfseries 1.0 \\
\cline{1-10}
\multirow[c]{4}{*}{DJI} & garch & $ -12.7 $ & $ -1.7 $ & - & - & - & - & - & - \\
 & rech & 3.7 & 0.3 & $ -1.5 $ & $ -0.4 $ & $ -3.6 $ & $ -0.5 $ & - & - \\
 & realgarch & 3.3 & 0.2 & $ -1.2 $ & $ -0.3 $ & - & - & $ -1.5 $ & $ -0.3 $ \\
 & deeprgarch  & \bfseries 5.5 & \bfseries 0.4 & \bfseries 1.6 & \bfseries 0.0 & \bfseries 2.6 & \bfseries 0.1 & \bfseries 0.3 & $ \mathbf{-0.1} $ \\
\cline{1-10}
\multirow[c]{4}{*}{GDAXI} & garch & $ -19.3 $ & $ -2.3 $ & - & - & - & - & - & - \\
 & rech & 9.6 & 0.9 & $ -1.0 $ & $ -0.2 $ & $ -4.7 $ & $ -0.5 $ & - & - \\
 & realgarch & $ -0.5 $ & $ -0.2 $ & $ -7.0 $ & $ -0.8 $ & - & - & $ -10.6 $ & $ -1.0 $ \\
 & deeprgarch  & \bfseries 11.7 & \bfseries 1.0 & \bfseries 7.0 & \bfseries 0.5 & \bfseries 3.3 & \bfseries 0.2 & \bfseries 10.0 & \bfseries 0.7 \\
\cline{1-10}
\multirow[c]{4}{*}{SPX} & garch & $ -17.4 $ & $ -2.4 $ & - & - & - & - & - & - \\
 & rech & 1.7 & $ -0.0 $ & $ -6.5 $ & $ -1.0 $ & $ -6.1 $ & $ -0.7 $ & - & - \\
 & realgarch & \bfseries 8.5 & \bfseries 0.9 & \bfseries 3.4 & \bfseries 0.2 & - & - & $ \mathbf{-0.0} $ & $ \mathbf{-0.1} $ \\
 & deeprgarch  & 8.4 & 0.8 & 2.3 & 0.1 & \bfseries 5.3 & \bfseries 0.4 & $ -1.1 $ & $ -0.2 $ \\
\cline{1-10}
\multirow[c]{4}{*}{Mean} & garch & $ -6.7 $ & $ -1.0 $ & - & - & - & - & - & - \\
 & rech & 5.2 & 0.4 & 2.0 & $ -0.0 $ & $ -2.8 $ & $ -0.4 $ & - & - \\
 & realgarch & $ -6.2 $ & $ -1.1 $ & $ -10.0 $ & $ -1.4 $ & - & - & $ -13.2 $ & $ -1.6 $ \\
 & deeprgarch  & \bfseries 10.8 & \bfseries 0.8 & \bfseries 11.1 & \bfseries 0.9 & \bfseries 4.1 & \bfseries 0.1 & \bfseries 19.1 & \bfseries 1.2 \\
\cline{1-10}
\multirow[c]{4}{*}{Count} & garch & 1 & 1 & 0 & 0 & 0 & 0 & 0 & 0 \\
 & rech & 8 & 9 & 9 & 9 & 9 & 9 & 0 & 0 \\
 & realgarch & 1 & 1 & 1 & 1 & 0 & 0 & 4 & 4 \\
 & deeprgarch  & \bfseries 21 & \bfseries 21 & \bfseries 21 & \bfseries 21 & \bfseries 22 & \bfseries 22 & \bfseries 27 & \bfseries 27 \\
\cline{1-10}
\bottomrule
\end{tabular}
\begin{tablenotes}
  \item \footnotesize \emph{Note:} Annualized returns are reported in cent(\%). The marks ``-" indicate models that are not trading. The last two panels report the average Ret./Sharpe and the number of times a model has the best predictive scores across 31 indices respectively. 
\end{tablenotes}
\end{threeparttable}
\end{table}
\end{spacing}

\subsection{Statistical significance}
The previous sections show that DeepRGARCH  outperforms the competing models in terms of in-sample fit, forecasting error, tail risk forecast and option pricing. This section tests whether these improvements are statistically significant using the Model Confidence Set (MCS) introduced by \cite{hansen_model_2011}. Let $\mathcal M$ be a set of competing models. A set of superior models (SSM) is established under the MCS procedure, which consists of a series of equal predictive accuracy tests given a specific confidence level. Let $L_{i,t}$ be a performance loss, such as the MSE or quantile loss, incurred by model $i\in \mathcal M$ at time $t$. Define $d_{i,j,t}=L_{i,t}-L_{j,t}$ to be the relative loss of model $i$ compared to model $j$ at time $t$.
The MCS test assumes that $d_{i,j,t}$ is a stationary time series for all $i,j$ in $\mathcal M$, i.e., $\mu_{i,j}=\E(d_{i,j,t})$ for all $t$. 
By testing the equality of the expected loss difference $\mu_{i,j}$, MCS determines if all models have the same level of predictive accuracy. The null hypothesis is 
\begin{equation}
    H_{0}: \mu_{i, j}=0, \quad \text{for all  } i, j \in \mathcal M.
\end{equation}
A model is eliminated when the null hypothesis $H_{0}$ of equal forecasting ability is rejected. The collection of models that do not reject the null hypothesis $H_{0}$ is then defined as the SSM. For each model $i\in\mathcal M$, the MCS produces a $p$-value $p_i$. The lower the $p$-value of a model, the less likely that it will be included in the SSM. See \cite{hansen_model_2011} for more details.\par

Table \ref{tab:mcs} reports the model confidence sets computed by all the predictive scores. For each model, we report the total number of times, across the 31 indices, that the model is included in the MCS and its average $p$-value (in parentheses). We note that a small $p$-value indicates that the model is unlikely to be the best model \parencite{hansen_model_2011}.
At a 75\% confidence level and across the predictive scores, the DeepRGARCH models are included in SSM for 24 indices, and have the highest $p$-value of 1 for 20 indices. The result shows that DeepRGARCH is the most likely to be included in the SSM and have statistically significant superiority over the other models in all the considered predictive metrics: forecasting error (MSE and MAD), fit to return series (PPS), risk forecast (quantile loss and joint loss) and option pricing. 

\begin{spacing}{}
\begin{table}[!htbp]
\centering
\begin{threeparttable}
\caption{Statistical signficance: The number of times a model is included in the set of superior models and the average $p$-value (in parentheses), across the 31 indices}
\label{tab:mcs}
\begin{tabular}{ccccc}
\toprule
 & GARCH & RECH & RealGARCH & DeepRGARCH \\
\midrule
MSE & \begin{tabular}[c]{@{}c@{}}1\\ {\footnotesize(0.01)}\end{tabular} & \begin{tabular}[c]{@{}c@{}}5\\ {\footnotesize(0.10)}\end{tabular} & \begin{tabular}[c]{@{}c@{}}18\\ {\footnotesize(0.48)}\end{tabular} & \begin{tabular}[c]{@{}c@{}}24\\ {\footnotesize(0.67)}\end{tabular} \\
\cline{1-5}
MAD & \begin{tabular}[c]{@{}c@{}}1\\ {\footnotesize(0.02)}\end{tabular} & \begin{tabular}[c]{@{}c@{}}4\\ {\footnotesize(0.09)}\end{tabular} & \begin{tabular}[c]{@{}c@{}}11\\ {\footnotesize(0.27)}\end{tabular} & \begin{tabular}[c]{@{}c@{}}26\\ {\footnotesize(0.84)}\end{tabular} \\
\cline{1-5}
Qloss\_1\% & \begin{tabular}[c]{@{}c@{}}9\\ {\footnotesize(0.16)}\end{tabular} & \begin{tabular}[c]{@{}c@{}}20\\ {\footnotesize(0.51)}\end{tabular} & \begin{tabular}[c]{@{}c@{}}12\\ {\footnotesize(0.25)}\end{tabular} & \begin{tabular}[c]{@{}c@{}}22\\ {\footnotesize(0.65)}\end{tabular} \\
\cline{1-5}
Qloss\_5\% & \begin{tabular}[c]{@{}c@{}}1\\ {\footnotesize(0.01)}\end{tabular} & \begin{tabular}[c]{@{}c@{}}12\\ {\footnotesize(0.30)}\end{tabular} & \begin{tabular}[c]{@{}c@{}}6\\ {\footnotesize(0.15)}\end{tabular} & \begin{tabular}[c]{@{}c@{}}25\\ {\footnotesize(0.80)}\end{tabular} \\
\cline{1-5}
JointLoss\_1\% & \begin{tabular}[c]{@{}c@{}}12\\ {\footnotesize(0.25)}\end{tabular} & \begin{tabular}[c]{@{}c@{}}16\\ {\footnotesize(0.40)}\end{tabular} & \begin{tabular}[c]{@{}c@{}}14\\ {\footnotesize(0.30)}\end{tabular} & \begin{tabular}[c]{@{}c@{}}24\\ {\footnotesize(0.70)}\end{tabular} \\
\cline{1-5}
JointLoss\_5\% & \begin{tabular}[c]{@{}c@{}}2\\ {\footnotesize(0.03)}\end{tabular} & \begin{tabular}[c]{@{}c@{}}13\\ {\footnotesize(0.31)}\end{tabular} & \begin{tabular}[c]{@{}c@{}}6\\ {\footnotesize(0.12)}\end{tabular} & \begin{tabular}[c]{@{}c@{}}25\\ {\footnotesize(0.80)}\end{tabular} \\
\cline{1-5}
PPS & \begin{tabular}[c]{@{}c@{}}7\\ {\footnotesize(0.13)}\end{tabular} & \begin{tabular}[c]{@{}c@{}}16\\ {\footnotesize(0.43)}\end{tabular} & \begin{tabular}[c]{@{}c@{}}10\\ {\footnotesize(0.24)}\end{tabular} & \begin{tabular}[c]{@{}c@{}}22\\ {\footnotesize(0.63)}\end{tabular} \\
\cline{1-5}
OptTrading & \begin{tabular}[c]{@{}c@{}}5\\ {\footnotesize(0.10)}\end{tabular} & \begin{tabular}[c]{@{}c@{}}13\\ {\footnotesize(0.39)}\end{tabular} & \begin{tabular}[c]{@{}c@{}}6\\ {\footnotesize(0.14)}\end{tabular} & \begin{tabular}[c]{@{}c@{}}22\\ {\footnotesize(0.70)}\end{tabular} \\
\cline{1-5}
\bottomrule
\end{tabular}
\end{threeparttable}
\end{table}
\end{spacing}

\section{Conclusions}\label{sec:conclusion}

The paper presents a novel methodology for volatility modeling, effectively harnessing the strengths of deep learning and the invaluable insights from realized volatility data. 
Bayesian inference and forecasting are performed using Sequential Monte Carlo with likelihood and data annealing. The DeepRGARCH model is evaluated on 31 major world stock markets, demonstrating improved performance in statistical criteria (in-sample fit, forecast error to realized measures) and economic criteria (tail risk forecast and option pricing) compared to the standard GARCH, RECH, and RealGARCH models.

Our proposed framework could be extended in several ways. First, it is possible to include multiple realized volatility measures since \cite{hansen_exponential_2016} discovered that including multiple realized volatility measures evidently improves both the in-sample and out-of-sample fit. The multiple measurement equations then could be replaced by a single LSTM with multiple outputs which would allow us to capture the non-linear relationship between conditional volatility and realized volatility measures as well as the interactions between the realized volatility measures. Second,
incorporating financial news into the input of DeepRGARCH is an interesting topic to study, as news has been shown to have a major influence on volatility movement (\cite{atkins_financial_2018}, \cite{xing_sentiment-aware_2019}, \cite{rahimikia_realised_2021}). Lastly, incorporating transfer learning would be a natural extension. Transfer learning can help mitigate the problem of insufficient training data and enable us to use a large number of pre-trained deep learning models and traditional financial econometric models for volatility modeling.

\newpage
\printbibliography

@article{liu_does_2015,
	title = {Does anything beat 5-minute {RV}? A comparison of realized measures across multiple asset classes},
	volume = {187},
	rights = {{RV}5 - Survey},
	shorttitle = {Does anything beat 5-minute {RV}?},
	pages = {293--311},
	number = {1},
	journaltitle = {Journal of Econometrics},
	author = {Liu, Lily Y. and Patton, Andrew J. and Sheppard, Kevin},
	date = {2015},
	note = {5mins {RV}},
}

@article{baillie1996fractionally,
  title={Fractionally integrated generalized autoregressive conditional heteroskedasticity},
  author={Baillie, Richard T and Bollerslev, Tim and Mikkelsen, Hans Ole},
  journal={Journal of econometrics},
  volume={74},
  number={1},
  pages={3--30},
  year={1996},
  publisher={Elsevier}
}

@article{Christensen_Nielsen_long_memory,
 ISSN = {00346535, 15309142},
 URL = {http://www.jstor.org/stable/40043094},
 author = {Bent Jesper Christensen and Morten Ørregaard Nielsen},
 journal = {The Review of Economics and Statistics},
 number = {4},
 pages = {684--700},
 publisher = {The MIT Press},
 title = {The Effect of Long Memory in Volatility on Stock Market Fluctuations},
 urldate = {2023-02-13},
 volume = {89},
 year = {2007}
}

@article{Benoit1967,
	ISSN = {00219398, 15375374},
	URL = {http://www.jstor.org/stable/2351623},
	author = {Benoit Mandelbrot},
	journal = {The Journal of Business},
	number = {4},
	pages = {393--413},
	publisher = {University of Chicago Press},
	title = {The Variation of Some Other Speculative Prices},
	volume = {40},
	year = {1967}
}

@article{Ding:1993,
	title = "A long memory property of stock market returns and a new model",
	journal = "Journal of Empirical Finance",
	volume = "1",
	number = "1",
	pages = "83 - 106",
	year = "1993",
	issn = "0927-5398",
	doi = "https://doi.org/10.1016/0927-5398(93)90006-D",
	url = "http://www.sciencedirect.com/science/article/pii/092753989390006D",
	author = "Zhuanxin Ding and Clive W.J. Granger and Robert F. Engle"
}

@article{atkins_financial_2018,
	title = {Financial news predicts stock market volatility better than close price},
	volume = {4},
	rights = {Later},
	pages = {120--137},
	number = {2},
	journaltitle = {The Journal of Finance and Data Science},
	author = {Atkins, Adam and Niranjan, Mahesan and Gerding, Enrico},
	date = {2018},
	note = {Publisher: Elsevier},
}

@article{bollerslev_generalized_1986,
	title = {Generalized autoregressive conditional heteroskedasticity},
	volume = {31},
	rights = {{GARCH}},
	issn = {0304-4076},
	url = {https://www.sciencedirect.com/science/article/pii/0304407686900631},
	doi = {10.1016/0304-4076(86)90063-1},
	pages = {307--327},
	number = {3},
	journaltitle = {Journal of Econometrics},
	shortjournal = {Journal of Econometrics},
	author = {Bollerslev, Tim},
	urldate = {2023-01-30},
	date = {1986-04-01},
	langid = {english},
}

@article{engle_autoregressive_1982,
	title = {Autoregressive Conditional Heteroscedasticity with Estimates of the Variance of United Kingdom Inflation},
	volume = {50},
	rights = {{ARCH}},
	issn = {0012-9682},
	url = {https://www.jstor.org/stable/1912773},
	doi = {10.2307/1912773},
	pages = {987--1007},
	number = {4},
	journaltitle = {Econometrica},
	author = {Engle, Robert F.},
	urldate = {2023-01-30},
	date = {1982},
	note = {Publisher: [Wiley, Econometric Society]},
}

@article{nelson_conditional_1991,
	title = {Conditional Heteroskedasticity in Asset Returns: A New Approach},
	volume = {59},
	rights = {{EGARCH}},
	issn = {0012-9682},
	url = {https://www.jstor.org/stable/2938260},
	doi = {10.2307/2938260},
	shorttitle = {Conditional Heteroskedasticity in Asset Returns},
	pages = {347--370},
	number = {2},
	journaltitle = {Econometrica},
	author = {Nelson, Daniel B.},
	urldate = {2023-01-30},
	date = {1991},
	note = {Publisher: [Wiley, Econometric Society]},
}

@article{glosten_relation_1993,
	title = {On the Relation between the Expected Value and the Volatility of the Nominal Excess Return on Stocks},
	volume = {48},
	rights = {{GJR}},
	issn = {1540-6261},
	url = {https://onlinelibrary.wiley.com/doi/abs/10.1111/j.1540-6261.1993.tb05128.x},
	doi = {10.1111/j.1540-6261.1993.tb05128.x},
	pages = {1779--1801},
	number = {5},
	journaltitle = {The Journal of Finance},
	author = {Glosten, Lawrence R. and Jagannathan, Ravi and Runkle, David E.},
	urldate = {2023-01-30},
	date = {1993},
	langid = {english},
	note = {\_eprint: https://onlinelibrary.wiley.com/doi/pdf/10.1111/j.1540-6261.1993.tb05128.x},
}

@article{forsberg_bridging_2002,
	title = {Bridging the gap between the distribution of realized ({ECU}) volatility and {ARCH} modelling (of the Euro): the {GARCH}-{NIG} model},
	volume = {17},
	rights = {{GARCHX}2},
	issn = {1099-1255},
	url = {https://onlinelibrary.wiley.com/doi/abs/10.1002/jae.685},
	doi = {10.1002/jae.685},
	shorttitle = {Bridging the gap between the distribution of realized ({ECU}) volatility and {ARCH} modelling (of the Euro)},
	pages = {535--548},
	number = {5},
	journaltitle = {Journal of Applied Econometrics},
	author = {Forsberg, Lars and Bollerslev, Tim},
	urldate = {2023-01-30},
	date = {2002},
	langid = {english},
	note = {\_eprint: https://onlinelibrary.wiley.com/doi/pdf/10.1002/jae.685},
}

@article{koenker_regression_1978,
	title = {Regression Quantiles},
	volume = {46},
	rights = {Qloss0},
	issn = {0012-9682},
	url = {https://www.jstor.org/stable/1913643},
	doi = {10.2307/1913643},
	pages = {33--50},
	number = {1},
	journaltitle = {Econometrica},
	author = {Koenker, Roger and Bassett, Gilbert},
	urldate = {2023-02-09},
	date = {1978},
	note = {Publisher: [Wiley, Econometric Society]},
}

@article{engle_new_2002,
	title = {New frontiers for arch models},
	volume = {17},
	rights = {{GARCHX}1},
	issn = {1099-1255},
	url = {https://onlinelibrary.wiley.com/doi/abs/10.1002/jae.683},
	doi = {10.1002/jae.683},
	pages = {425--446},
	number = {5},
	journaltitle = {Journal of Applied Econometrics},
	author = {Engle, Robert},
	urldate = {2023-01-30},
	date = {2002},
	langid = {english},
	note = {\_eprint: https://onlinelibrary.wiley.com/doi/pdf/10.1002/jae.683},
}

@article{engle_multiple_2006,
	title = {A multiple indicators model for volatility using intra-daily data},
	volume = {131},
	rights = {{MEM}},
	issn = {0304-4076},
	url = {https://www.sciencedirect.com/science/article/pii/S0304407605000047},
	doi = {10.1016/j.jeconom.2005.01.018},
	pages = {3--27},
	number = {1},
	journaltitle = {Journal of Econometrics},
	shortjournal = {Journal of Econometrics},
	author = {Engle, Robert F. and Gallo, Giampiero M.},
	urldate = {2023-01-30},
	date = {2006-03-01},
	langid = {english},
}

@article{corsi_simple_2009,
	title = {A Simple Approximate Long-Memory Model of Realized Volatility},
	volume = {7},
	rights = {{HAR}},
	issn = {1479-8409},
	url = {https://doi.org/10.1093/jjfinec/nbp001},
	doi = {10.1093/jjfinec/nbp001},
	pages = {174--196},
	number = {2},
	journaltitle = {Journal of Financial Econometrics},
	shortjournal = {Journal of Financial Econometrics},
	author = {Corsi, Fulvio},
	urldate = {2023-01-30},
	date = {2009-03-01},
}

@article{shephard_realising_2010,
	title = {Realising the future: forecasting with high-frequency-based volatility ({HEAVY}) models},
	volume = {25},
	rights = {{MEM}2},
	issn = {1099-1255},
	url = {https://onlinelibrary.wiley.com/doi/abs/10.1002/jae.1158},
	doi = {10.1002/jae.1158},
	shorttitle = {Realising the future},
	pages = {197--231},
	number = {2},
	journaltitle = {Journal of Applied Econometrics},
	author = {Shephard, Neil and Sheppard, Kevin},
	urldate = {2023-01-30},
	date = {2010},
	langid = {english},
	note = {\_eprint: https://onlinelibrary.wiley.com/doi/pdf/10.1002/jae.1158},
}

@article{hansen_realized_2012,
	title = {Realized {GARCH}: a joint model for returns and realized measures of volatility},
	volume = {27},
	rights = {{RealGARCH}},
	issn = {1099-1255},
	url = {https://onlinelibrary.wiley.com/doi/abs/10.1002/jae.1234},
	doi = {10.1002/jae.1234},
	shorttitle = {Realized {GARCH}},
	pages = {877--906},
	number = {6},
	journaltitle = {Journal of Applied Econometrics},
	author = {Hansen, Peter Reinhard and Huang, Zhuo and Shek, Howard Howan},
	urldate = {2023-01-30},
	date = {2012},
	langid = {english},
	note = {\_eprint: https://onlinelibrary.wiley.com/doi/pdf/10.1002/jae.1234},
}

@article{hansen_exponential_2016,
	title = {Exponential {GARCH} Modeling With Realized Measures of Volatility},
	volume = {34},
	rights = {{RealEGARCH}},
	issn = {0735-0015},
	url = {https://doi.org/10.1080/07350015.2015.1038543},
	doi = {10.1080/07350015.2015.1038543},
	pages = {269--287},
	number = {2},
	journaltitle = {Journal of Business \& Economic Statistics},
	author = {Hansen, Peter Reinhard and Huang, Zhuo},
	urldate = {2023-01-30},
	date = {2016-04-02},
	note = {Publisher: Taylor \& Francis
\_eprint: https://doi.org/10.1080/07350015.2015.1038543},
}

@article{xing_sentiment-aware_2019,
	title = {Sentiment-aware volatility forecasting},
	volume = {176},
	rights = {News},
	issn = {0950-7051},
	url = {https://www.sciencedirect.com/science/article/pii/S0950705119301546},
	doi = {10.1016/j.knosys.2019.03.029},
	pages = {68--76},
	journaltitle = {Knowledge-Based Systems},
	shortjournal = {Knowledge-Based Systems},
	author = {Xing, Frank Z. and Cambria, Erik and Zhang, Yue},
	urldate = {2023-01-30},
	date = {2019-07-15},
	langid = {english},
}

@article{rahimikia_realised_2021,
	title = {Realised Volatility Forecasting: Machine Learning via Financial Word Embedding},
	rights = {{NN} - news},
	issn = {1556-5068},
	url = {http://arxiv.org/abs/2108.00480},
	doi = {10.2139/ssrn.3895272},
	shorttitle = {Realised Volatility Forecasting},
	journaltitle = {{SSRN} Electronic Journal},
	shortjournal = {{SSRN} Journal},
	author = {Rahimikia, Eghbal and Zohren, Stefan and Poon, Ser-Huang},
	urldate = {2023-01-30},
	date = {2021},
	eprinttype = {arxiv},
	eprint = {2108.00480 [cs, q-fin]},
}

@article{nguyen_recurrent_2022,
	title = {Recurrent conditional heteroskedasticity},
	volume = {37},
	rights = {{RECH}},
	issn = {1099-1255},
	url = {https://onlinelibrary.wiley.com/doi/abs/10.1002/jae.2902},
	doi = {10.1002/jae.2902},
	pages = {1031--1054},
	number = {5},
	journaltitle = {Journal of Applied Econometrics},
	author = {Nguyen, Trong-Nghia and Tran, Minh-Ngoc and Kohn, Robert},
	urldate = {2023-01-30},
	date = {2022},
	langid = {english},
	note = {\_eprint: https://onlinelibrary.wiley.com/doi/pdf/10.1002/jae.2902},
}

@article{andersen_answering_1998,
	title = {Answering the Skeptics: Yes, Standard Volatility Models do Provide Accurate Forecasts},
	volume = {39},
	rights = {Evaluate {GARCH}},
	issn = {0020-6598},
	url = {https://www.jstor.org/stable/2527343},
	doi = {10.2307/2527343},
	shorttitle = {Answering the Skeptics},
	pages = {885--905},
	number = {4},
	journaltitle = {International Economic Review},
	author = {Andersen, Torben G. and Bollerslev, Tim},
	urldate = {2023-02-05},
	date = {1998},
	note = {Publisher: [Economics Department of the University of Pennsylvania, Wiley, Institute of Social and Economic Research, Osaka University]},
}

@article{andersen_modeling_2003,
	title = {Modeling and Forecasting Realized Volatility},
	volume = {71},
	rights = {{RV} - overall},
	issn = {1468-0262},
	url = {https://onlinelibrary.wiley.com/doi/abs/10.1111/1468-0262.00418},
	doi = {10.1111/1468-0262.00418},
	pages = {579--625},
	number = {2},
	journaltitle = {Econometrica},
	author = {Andersen, Torben G. and Bollerslev, Tim and Diebold, Francis X. and Labys, Paul},
	urldate = {2023-02-05},
	date = {2003},
	langid = {english},
	note = {\_eprint: https://onlinelibrary.wiley.com/doi/pdf/10.1111/1468-0262.00418},
}

@article{barndorff-nielsen_power_2004,
	title = {Power and Bipower Variation with Stochastic Volatility and Jumps},
	volume = {2},
	rights = {{BV}},
	issn = {1479-8409},
	url = {https://doi.org/10.1093/jjfinec/nbh001},
	doi = {10.1093/jjfinec/nbh001},
	pages = {1--37},
	number = {1},
	journaltitle = {Journal of Financial Econometrics},
	shortjournal = {Journal of Financial Econometrics},
	author = {Barndorff-Nielsen, Ole E. and Shephard, Neil},
	urldate = {2023-02-05},
	date = {2004-01-01},
}

@article{barndorff-nielsen_designing_2008,
	title = {Designing Realized Kernels to Measure the ex post Variation of Equity Prices in the Presence of Noise},
	volume = {76},
	rights = {{RKV} - Non-Flat Parzen kernal},
	issn = {1468-0262},
	url = {https://onlinelibrary.wiley.com/doi/abs/10.3982/ECTA6495},
	doi = {10.3982/ECTA6495},
	pages = {1481--1536},
	number = {6},
	journaltitle = {Econometrica},
	author = {Barndorff-Nielsen, Ole E. and Hansen, Peter Reinhard and Lunde, Asger and Shephard, Neil},
	urldate = {2023-02-05},
	date = {2008},
	langid = {english},
	note = {\_eprint: https://onlinelibrary.wiley.com/doi/pdf/10.3982/{ECTA}6495},
}

@article{bucci_realized_2020,
	title = {Realized Volatility Forecasting with Neural Networks},
	volume = {18},
	rights = {{LSTM} - rvs},
	issn = {1479-8409},
	url = {https://doi.org/10.1093/jjfinec/nbaa008},
	doi = {10.1093/jjfinec/nbaa008},
	pages = {502--531},
	number = {3},
	journaltitle = {Journal of Financial Econometrics},
	shortjournal = {Journal of Financial Econometrics},
	author = {Bucci, Andrea},
	urldate = {2023-02-05},
	date = {2020-06-01},
}

@article{chopin_sequential_2002,
	title = {A sequential particle filter method for static models},
	volume = {89},
	rights = {Data annealing},
	issn = {0006-3444},
	url = {https://doi.org/10.1093/biomet/89.3.539},
	doi = {10.1093/biomet/89.3.539},
	pages = {539--552},
	number = {3},
	journaltitle = {Biometrika},
	shortjournal = {Biometrika},
	author = {Chopin, Nicolas},
	urldate = {2023-02-05},
	date = {2002-08-01},
}

@article{del_moral_adaptive_2012,
	title = {An adaptive sequential Monte Carlo method for approximate Bayesian computation},
	volume = {22},
	rights = {Adaptivve tempering},
	issn = {1573-1375},
	url = {https://doi.org/10.1007/s11222-011-9271-y},
	doi = {10.1007/s11222-011-9271-y},
	pages = {1009--1020},
	number = {5},
	journaltitle = {Statistics and Computing},
	shortjournal = {Stat Comput},
	author = {Del Moral, Pierre and Doucet, Arnaud and Jasra, Ajay},
	urldate = {2023-02-05},
	date = {2012-09-01},
	langid = {english},
}

@article{hansen_forecast_2005,
	title = {A forecast comparison of volatility models: does anything beat a {GARCH}(1,1)?},
	volume = {20},
	rights = {{GARCH}(1,1)},
	issn = {1099-1255},
	url = {https://onlinelibrary.wiley.com/doi/abs/10.1002/jae.800},
	doi = {10.1002/jae.800},
	shorttitle = {A forecast comparison of volatility models},
	pages = {873--889},
	number = {7},
	journaltitle = {Journal of Applied Econometrics},
	author = {Hansen, Peter R. and Lunde, Asger},
	urldate = {2023-02-05},
	date = {2005},
	langid = {english},
	note = {\_eprint: https://onlinelibrary.wiley.com/doi/pdf/10.1002/jae.800},
}

@article{hansen_model_2011,
	title = {The Model Confidence Set},
	volume = {79},
	rights = {{MCS}},
	issn = {1468-0262},
	url = {https://onlinelibrary.wiley.com/doi/abs/10.3982/ECTA5771},
	doi = {10.3982/ECTA5771},
	pages = {453--497},
	number = {2},
	journaltitle = {Econometrica},
	author = {Hansen, Peter R. and Lunde, Asger and Nason, James M.},
	urldate = {2023-02-05},
	date = {2011},
	langid = {english},
	note = {\_eprint: https://onlinelibrary.wiley.com/doi/pdf/10.3982/{ECTA}5771},
}

@article{hochreiter_long_1997,
	title = {Long Short-Term Memory},
	volume = {9},
	rights = {{LSTM}},
	issn = {0899-7667},
	doi = {10.1162/neco.1997.9.8.1735},
	pages = {1735--1780},
	number = {8},
	journaltitle = {Neural Computation},
	author = {Hochreiter, Sepp and Schmidhuber, Jürgen},
	date = {1997-11},
	note = {Conference Name: Neural Computation},
}

@book{jeffreys_theory_1998,
	title = {The Theory of Probability},
	rights = {Probability},
	isbn = {978-0-19-158967-6},
	pagetotal = {474},
	publisher = {{OUP} Oxford},
	author = {Jeffreys, Harold},
	date = {1998-08-06},
	langid = {english},
	note = {Google-Books-{ID}: vh9Act9rtzQC},
}

@article{jiang_modeling_2018,
	title = {Modeling returns volatility: Realized {GARCH} incorporating realized risk measure},
	volume = {500},
	rights = {Later},
	issn = {0378-4371},
	url = {https://www.sciencedirect.com/science/article/pii/S0378437118300943},
	doi = {10.1016/j.physa.2018.02.018},
	shorttitle = {Modeling returns volatility},
	pages = {249--258},
	journaltitle = {Physica A: Statistical Mechanics and its Applications},
	shortjournal = {Physica A: Statistical Mechanics and its Applications},
	author = {Jiang, Wei and Ruan, Qingsong and Li, Jianfeng and Li, Ye},
	urldate = {2023-02-05},
	date = {2018-06-15},
	langid = {english},
}

@article{kass_bayes_1995,
	title = {Bayes Factors},
	volume = {90},
	rights = {Bayes factors},
	issn = {0162-1459},
	url = {https://www.tandfonline.com/doi/abs/10.1080/01621459.1995.10476572},
	doi = {10.1080/01621459.1995.10476572},
	pages = {773--795},
	number = {430},
	journaltitle = {Journal of the American Statistical Association},
	author = {Kass, Robert E. and Raftery, Adrian E.},
	urldate = {2023-02-05},
	date = {1995-06-01},
	note = {Publisher: Taylor \& Francis
\_eprint: https://www.tandfonline.com/doi/pdf/10.1080/01621459.1995.10476572},
}

@article{kass_markov_1998,
	title = {Markov Chain Monte Carlo in Practice: A Roundtable Discussion},
	volume = {52},
	rights = {Later},
	issn = {0003-1305},
	url = {https://www.tandfonline.com/doi/abs/10.1080/00031305.1998.10480547},
	doi = {10.1080/00031305.1998.10480547},
	shorttitle = {Markov Chain Monte Carlo in Practice},
	pages = {93--100},
	number = {2},
	journaltitle = {The American Statistician},
	author = {Kass, Robert E. and Carlin, Bradley P. and Gelman, Andrew and Neal, Radford M.},
	urldate = {2023-02-05},
	date = {1998-05-01},
	note = {Publisher: Taylor \& Francis
\_eprint: https://www.tandfonline.com/doi/pdf/10.1080/00031305.1998.10480547},
}

@article{kim_forecasting_2018,
	title = {Forecasting the volatility of stock price index: A hybrid model integrating {LSTM} with multiple {GARCH}-type models},
	volume = {103},
	rights = {Later},
	issn = {0957-4174},
	url = {https://www.sciencedirect.com/science/article/pii/S0957417418301416},
	doi = {10.1016/j.eswa.2018.03.002},
	shorttitle = {Forecasting the volatility of stock price index},
	pages = {25--37},
	journaltitle = {Expert Systems with Applications},
	shortjournal = {Expert Systems with Applications},
	author = {Kim, Ha Young and Won, Chang Hyun},
	urldate = {2023-02-05},
	date = {2018-08-01},
	langid = {english},
}

@article{li_efficient_2021,
	title = {Efficient Bayesian estimation for {GARCH}-type models via Sequential Monte Carlo},
	volume = {19},
	rights = {Later},
	issn = {2452-3062},
	url = {https://www.sciencedirect.com/science/article/pii/S2452306220300319},
	doi = {10.1016/j.ecosta.2020.02.002},
	pages = {22--46},
	journaltitle = {Econometrics and Statistics},
	shortjournal = {Econometrics and Statistics},
	author = {Li, Dan and Clements, Adam and Drovandi, Christopher},
	urldate = {2023-02-05},
	date = {2021-07-01},
	langid = {english},
}

@article{neal_annealed_2001,
	title = {Annealed importance sampling},
	volume = {11},
	rights = {Likelihood annealing},
	issn = {1573-1375},
	url = {https://doi.org/10.1023/A:1008923215028},
	doi = {10.1023/A:1008923215028},
	pages = {125--139},
	number = {2},
	journaltitle = {Statistics and Computing},
	shortjournal = {Statistics and Computing},
	author = {Neal, Radford M.},
	urldate = {2023-02-05},
	date = {2001-04-01},
	langid = {english},
}

@article{hyup_roh_forecasting_2007,
	title = {Forecasting the volatility of stock price index},
	volume = {33},
	rights = {Later},
	issn = {0957-4174},
	url = {https://www.sciencedirect.com/science/article/pii/S0957417406002223},
	doi = {10.1016/j.eswa.2006.08.001},
	pages = {916--922},
	number = {4},
	journaltitle = {Expert Systems with Applications},
	shortjournal = {Expert Systems with Applications},
	author = {Hyup Roh, Tae},
	urldate = {2023-02-05},
	date = {2007-11-01},
	langid = {english},
}

@article{xie_realized_2020,
	title = {Realized {GARCH} models: Simpler is better},
	volume = {33},
	rights = {Later},
	issn = {1544-6123},
	url = {https://www.sciencedirect.com/science/article/pii/S1544612318308365},
	doi = {10.1016/j.frl.2019.06.019},
	shorttitle = {Realized {GARCH} models},
	pages = {101221},
	journaltitle = {Finance Research Letters},
	shortjournal = {Finance Research Letters},
	author = {Xie, Haibin and Yu, Chengtan},
	urldate = {2023-02-05},
	date = {2020-03-01},
	langid = {english},
}

@article{gunawan_flexible_2022,
	title = {Flexible and Robust Particle Tempering for State Space Models},
	rights = {Later},
	issn = {2452-3062},
	url = {https://www.sciencedirect.com/science/article/pii/S2452306222000843},
	doi = {10.1016/j.ecosta.2022.09.003},
	journaltitle = {Econometrics and Statistics},
	shortjournal = {Econometrics and Statistics},
	author = {Gunawan, David and Kohn, Robert and Tran, Minh Ngoc},
	urldate = {2023-02-05},
	date = {2022-10-10},
	langid = {english},
}

@article{gerlach_forecasting_2016,
	title = {Forecasting risk via realized {GARCH}, incorporating the realized range},
	volume = {16},
	rights = {Later},
	issn = {1469-7688},
	url = {https://doi.org/10.1080/14697688.2015.1079641},
	doi = {10.1080/14697688.2015.1079641},
	pages = {501--511},
	number = {4},
	journaltitle = {Quantitative Finance},
	author = {Gerlach, Richard and Wang, Chao},
	urldate = {2023-02-05},
	date = {2016-04-02},
	note = {Publisher: Routledge
\_eprint: https://doi.org/10.1080/14697688.2015.1079641},
}

@article{fissler_higher_2016,
	title = {Higher order elicitability and Osband’s principle},
	volume = {44},
	rights = {Later},
	issn = {0090-5364, 2168-8966},
	url = {https://projecteuclid.org/journals/annals-of-statistics/volume-44/issue-4/Higher-order-elicitability-and-Osbands-principle/10.1214/16-AOS1439.full},
	doi = {10.1214/16-AOS1439},
	pages = {1680--1707},
	number = {4},
	journaltitle = {The Annals of Statistics},
	author = {Fissler, Tobias and Ziegel, Johanna F.},
	urldate = {2023-02-05},
	date = {2016-08},
	note = {Publisher: Institute of Mathematical Statistics},
}

@misc{engle_valuation_1990,
	title = {Valuation of Variance Forecast with Simulated Option Markets},
	rights = {Option simulation},
	url = {https://www.nber.org/papers/w3350},
	doi = {10.3386/w3350},
	series = {Working Paper Series},
	number = {3350},
	publisher = {National Bureau of Economic Research},
	author = {Engle, Robert F. and Hong, Che-Hsiung and Kane, Alex},
	urldate = {2023-02-05},
	date = {1990-05},
	doi = {10.3386/w3350},
}

@article{liu_novel_2019,
	title = {Novel volatility forecasting using deep learning–long short term memory recurrent neural networks},
	volume = {132},
	rights = {Later},
	pages = {99--109},
	journaltitle = {Expert Systems with Applications},
	author = {Liu, Yang},
	date = {2019},
	note = {Publisher: Elsevier},
}

@article{taylor_forecasting_2019,
	title = {Forecasting Value at Risk and Expected Shortfall Using a Semiparametric Approach Based on the Asymmetric Laplace Distribution},
	volume = {37},
	rights = {Later},
	issn = {0735-0015},
	url = {https://doi.org/10.1080/07350015.2017.1281815},
	doi = {10.1080/07350015.2017.1281815},
	pages = {121--133},
	number = {1},
	journaltitle = {Journal of Business \& Economic Statistics},
	author = {Taylor, James W.},
	urldate = {2023-02-05},
	date = {2019-01-02},
	note = {Publisher: Taylor \& Francis
\_eprint: https://doi.org/10.1080/07350015.2017.1281815},
}

@book{goodfellow_deep_2016,
	title = {Deep Learning},
	rights = {Deep learning},
	isbn = {978-0-262-33737-3},
	pagetotal = {801},
	publisher = {{MIT} Press},
	author = {Goodfellow, Ian and Bengio, Yoshua and Courville, Aaron},
	date = {2016-11-10},
	langid = {english},
	note = {Google-Books-{ID}: {omivDQAAQBAJ}},
}

@article{bollerslev_glossary_2008,
	title = {Glossary to {ARCH} ({GARCH})},
	volume = {2008},
	rights = {Later},
	issn = {1556-5068},
	url = {http://www.ssrn.com/abstract=1263250},
	doi = {10.2139/ssrn.1263250},
	number = {49},
	journaltitle = {{CREATES} Research Paper},
	shortjournal = {{SSRN} Journal},
	author = {Bollerslev, Tim},
	urldate = {2023-02-05},
	date = {2008},
	langid = {english},
}
\newpage
\singlespacing
\section{Appendix}

\begin{spacing}{}
\begin{longtable}[!htbp]{ccccc}
\caption{Datasets specification (all indices).} \label{datafull} \\
\toprule
 & In-Sample & Out-of-Sample & T\_in & T\_out \\
\midrule
\endfirsthead
\bottomrule
\endlastfoot
AEX & 2004-01-02 - 2012-12-21 & 2012-12-24 - 2021-12-31 & 2302 & 2302 \\
AORD & 2004-01-02 - 2013-01-02 & 2013-01-03 - 2021-12-31 & 2276 & 2277 \\
BFX & 2004-01-02 - 2012-12-21 & 2012-12-24 - 2021-12-31 & 2302 & 2302 \\
BSESN & 2004-01-01 - 2012-12-20 & 2012-12-21 - 2021-12-31 & 2233 & 2234 \\
BVLG & 2012-10-15 - 2017-05-18 & 2017-05-21 - 2021-12-31 & 1171 & 1172 \\
BVSP & 2004-01-02 - 2012-12-17 & 2012-12-18 - 2021-12-30 & 2216 & 2216 \\
DJI & 2004-01-02 - 2012-12-26 & 2012-12-27 - 2021-12-31 & 2258 & 2258 \\
FCHI & 2004-01-02 - 2012-12-24 & 2012-12-27 - 2021-12-31 & 2303 & 2304 \\
\end{longtable}

\end{spacing}

\begin{spacing}{}
\begin{longtable}[!htbp]{ccccccccc}
\caption{In-sample Analysis: Posterior means of the parameters with the posterior standard deviations (all indices)} \label{parafull} \\
\toprule
 &  & $\alpha$ & $\beta$ & $\beta_0$ & $\beta_1$ & $\gamma$ & MargLik & BF \\
\midrule
\endfirsthead
\bottomrule
\endlastfoot
\multirow[c]{4}{*}{AEX} & garch & \begin{tabular}[c]{@{}c@{}}0.105\\ {\scriptsize(0.012)}\end{tabular} & \begin{tabular}[c]{@{}c@{}}0.884\\ {\scriptsize(0.013)}\end{tabular} & - & - & - & $ -3450.8 $ & 1 \\
 & rech & \begin{tabular}[c]{@{}c@{}}0.038\\ {\scriptsize(0.014)}\end{tabular} & \begin{tabular}[c]{@{}c@{}}0.741\\ {\scriptsize(0.061)}\end{tabular} & \begin{tabular}[c]{@{}c@{}}0.049\\ {\scriptsize(0.019)}\end{tabular} & \begin{tabular}[c]{@{}c@{}}0.907\\ {\scriptsize(0.182)}\end{tabular} & - & $ -3400.6 $ & $e^{50}$ \\
 & realgarch & - & \begin{tabular}[c]{@{}c@{}}0.549\\ {\scriptsize(0.028)}\end{tabular} & - & - & \begin{tabular}[c]{@{}c@{}}0.371\\ {\scriptsize(0.028)}\end{tabular} & $ -3376.3 $ & $e^{75}$ \\
 & deeprgarch  & - & \begin{tabular}[c]{@{}c@{}}0.527\\ {\scriptsize(0.021)}\end{tabular} & \begin{tabular}[c]{@{}c@{}}0.112\\ {\scriptsize(0.022)}\end{tabular} & \begin{tabular}[c]{@{}c@{}}0.801\\ {\scriptsize(0.081)}\end{tabular} & \begin{tabular}[c]{@{}c@{}}0.341\\ {\scriptsize(0.015)}\end{tabular} & $ \mathbf{-3370.5} $ & $\mathbf{e^{80}}$ \\
\cline{1-9}
\multirow[c]{4}{*}{AORD} & garch & \begin{tabular}[c]{@{}c@{}}0.101\\ {\scriptsize(0.012)}\end{tabular} & \begin{tabular}[c]{@{}c@{}}0.894\\ {\scriptsize(0.013)}\end{tabular} & - & - & - & $ -2911.1 $ & 1 \\
 & rech & \begin{tabular}[c]{@{}c@{}}0.055\\ {\scriptsize(0.014)}\end{tabular} & \begin{tabular}[c]{@{}c@{}}0.792\\ {\scriptsize(0.040)}\end{tabular} & \begin{tabular}[c]{@{}c@{}}0.017\\ {\scriptsize(0.008)}\end{tabular} & \begin{tabular}[c]{@{}c@{}}0.736\\ {\scriptsize(0.181)}\end{tabular} & - & $ -2883.7 $ & $e^{27}$ \\
 & realgarch & - & \begin{tabular}[c]{@{}c@{}}0.744\\ {\scriptsize(0.020)}\end{tabular} & - & - & \begin{tabular}[c]{@{}c@{}}0.199\\ {\scriptsize(0.017)}\end{tabular} & $ -2886.9 $ & $e^{24}$ \\
\end{longtable}
\end{spacing}

\begin{spacing}{}
\begin{longtable}[!htbp]{cccccccc}
\caption{Forecast performance: MSE for different realized volatility measures (all indices)} \label{msefull} \\
\toprule
 &  & rv5 & bv & medrv & rk\_parzen & rsv & count \\
\midrule
\endfirsthead
\bottomrule
\endlastfoot
\multirow[c]{4}{*}{AEX} & garch & 0.151 & 0.142 & 0.141 & 0.209 & 0.190 & 0.0 \\
 & rech & 0.143 & 0.131 & 0.126 & 0.188 & 0.180 & 0.0 \\
 & realgarch & 0.121 & 0.109 & 0.102 & 0.173 & 0.158 & 0.0 \\
 & deeprgarch  & \bfseries 0.116 & \bfseries 0.103 & \bfseries 0.094 & \bfseries 0.157 & \bfseries 0.152 & \bfseries 5.0 \\
\cline{1-8}
\multirow[c]{4}{*}{AORD} & garch & 0.169 & 0.169 & 0.109 & 0.241 & 0.299 & 0.0 \\
 & rech & 0.170 & 0.171 & 0.115 & 0.239 & 0.301 & 0.0 \\
 & realgarch & 0.149 & 0.150 & 0.100 & 0.228 & 0.292 & 0.0 \\
 & deeprgarch  & \bfseries 0.141 & \bfseries 0.142 & \bfseries 0.093 & \bfseries 0.215 & \bfseries 0.275 & \bfseries 5.0 \\
\end{longtable}
\end{spacing}

\begin{spacing}{}
\begin{longtable}[!htbp]{cccccccc}
\caption{Forecast performance: MAD for different realized volatility measures (all indices).} \label{madfull} \\
\toprule
 &  & rv5 & bv & medrv & rk\_parzen & rsv & count \\
\midrule
\endfirsthead
\bottomrule
\endlastfoot
\multirow[c]{4}{*}{AEX} & garch & 0.259 & 0.256 & 0.266 & 0.332 & 0.298 & 0.0 \\
 & rech & 0.237 & 0.233 & 0.242 & 0.300 & 0.276 & 0.0 \\
 & realgarch & 0.218 & 0.211 & 0.219 & 0.294 & 0.260 & 0.0 \\
 & deeprgarch  & \bfseries 0.209 & \bfseries 0.202 & \bfseries 0.209 & \bfseries 0.274 & \bfseries 0.249 & \bfseries 5.0 \\
\cline{1-8}
\multirow[c]{4}{*}{AORD} & garch & 0.231 & 0.232 & 0.197 & 0.309 & 0.318 & 0.0 \\
 & rech & 0.222 & 0.224 & 0.197 & \bfseries 0.298 & \bfseries 0.312 & 2.0 \\
 & realgarch & 0.229 & 0.231 & 0.203 & 0.315 & 0.336 & 0.0 \\
 & deeprgarch  & \bfseries 0.218 & \bfseries 0.220 & \bfseries 0.194 & 0.301 & 0.316 & \bfseries 3.0 \\
\end{longtable}
\end{spacing}

\begin{spacing}{}
\begin{longtable}[!htbp]{ccccccc}
\caption{Forecast performance: Tail risk forecast and Partial Predictive Score (all indices).} \label{riskfull} \\
\toprule
 &  & Qloss\_1\% & JointLoss\_1\% & Qloss\_5\% & JointLoss\_5\% & PPS \\
\midrule
\endfirsthead
\bottomrule
\endlastfoot
\multirow[c]{4}{*}{AEX} & garch & 92.549 & 2.462 & 282.507 & 1.909 & 1.342 \\
 & rech & 88.654 & 2.404 & 270.841 & 1.851 & 1.307 \\
 & realgarch & 86.838 & 2.317 & 273.663 & 1.847 & 1.298 \\
 & deeprgarch  & \bfseries 84.982 & \bfseries 2.279 & \bfseries 267.902 & \bfseries 1.815 & \bfseries 1.288 \\
\cline{1-7}
\multirow[c]{4}{*}{AORD} & garch & 46.639 & \bfseries 1.643 & 191.509 & 1.467 & 1.059 \\
 & rech & 46.556 & 1.649 & 183.081 & 1.435 & \bfseries 1.054 \\
 & realgarch & 51.403 & 1.758 & 198.293 & 1.528 & 1.095 \\
 & deeprgarch  & \bfseries 44.988 & 1.664 & \bfseries 172.360 & \bfseries 1.422 & 1.058 \\
\end{longtable}
\end{spacing}

\begin{spacing}{}
\begin{longtable}[!htbp]{rrrrrrrrrr}
\caption{Forecast performance: Annualized return (Ret.) and Sharpe ratio (Sharpe) of option trading simulation (all indices)} \label{optfull} \\
\toprule
 &  & \multicolumn{2}{r}{Scenario1} & \multicolumn{2}{r}{Scenario2} & \multicolumn{2}{r}{Scenario3} & \multicolumn{2}{r}{Scenario4} \\
 &  & Ret. & Sharpe & Ret. & Sharpe & Ret. & Sharpe & Ret. & Sharpe \\
\midrule
\endfirsthead
\bottomrule
\endlastfoot
\multirow[c]{4}{*}{AEX} & garch & $ -22.4 $ & $ -3.2 $ & - & - & - & - & - & - \\
 & rech & 5.9 & 0.5 & $ -4.1 $ & $ -0.6 $ & $ -9.5 $ & $ -1.1 $ & - & - \\
 & realgarch & 1.9 & 0.0 & $ -7.4 $ & $ -1.0 $ & - & - & $ -12.4 $ & $ -1.4 $ \\
 & deeprgarch  & \bfseries 17.8 & \bfseries 1.9 & \bfseries 11.3 & \bfseries 1.2 & \bfseries 9.2 & \bfseries 0.7 & \bfseries 12.9 & \bfseries 1.0 \\
\cline{1-10}
\multirow[c]{4}{*}{AORD} & garch & 1.9 & 0.0 & - & - & - & - & - & - \\
 & rech & 4.7 & 0.5 & 5.4 & 0.5 & $ -6.3 $ & $ -1.0 $ & - & - \\
 & realgarch & $ -19.8 $ & $ -3.9 $ & $ -22.8 $ & $ -4.1 $ & - & - & $ -29.2 $ & $ -4.7 $ \\
 & deeprgarch  & \bfseries 16.1 & \bfseries 2.1 & \bfseries 22.1 & \bfseries 2.9 & \bfseries 6.0 & \bfseries 0.5 & \bfseries 40.3 & \bfseries 4.2 \\
\end{longtable}
\end{spacing}

\end{document}